\documentclass[sigconf, nonacm]{acmart}

\AtBeginDocument{%
  }

\setcopyright{none} 
\copyrightyear{2026}
\acmYear{2026}
\acmDOI{XXXXXXX.XXXXXXX}

\acmConference[MICRO 2026]{The 58th IEEE/ACM International Symposium on Microarchitecture}{October 31--November 04, 2026}{Athens, Greece}

\acmISBN{978-X-XXXX-XXXX-X/XX/XX}

\usepackage{amsmath,amssymb,amsfonts}
\usepackage{algorithm}          
\usepackage{algpseudocode} 
\usepackage{graphicx}
\usepackage[normalem]{ulem}
\usepackage{textcomp}
\usepackage{xcolor}
\usepackage{multirow}
\usepackage{booktabs}
\usepackage[caption=false,font=footnotesize]{subfig}
\usepackage{makecell}
\usepackage{pifont}
\usepackage{url}
\usepackage{soul}  
\usepackage{colortbl} 
\usepackage{threeparttable}
\usepackage{color}
\usepackage[table]{xcolor}

\begin{document}
\begin{sloppypar}
\title{VersaQ-3D: Architecture Support for Visual Geometry Grounded Transformers via Versatile Quantization}
\title{VersaQ-3D: Architecture Support for Visual Geometry Grounded Transformers via Versatile Quantization}

\author{Yipu Zhang\textsuperscript{1},
Jintao Cheng\textsuperscript{1},
Xingyu Liu\textsuperscript{1},
Zeyu Li\textsuperscript{1},
Carol Jingyi Li\textsuperscript{1},
Jin Wu\textsuperscript{2},
Lin Jiang\textsuperscript{3},
Ceyu Xu\textsuperscript{1},
Yuan Xie\textsuperscript{1},
Jiang Xu\textsuperscript{4},
Wei Zhang\textsuperscript{1,*}}

\affiliation{%
  \institution{%
    \textsuperscript{1}HKUST,
    \textsuperscript{2}USTB,
    \textsuperscript{3}NEU,
    \textsuperscript{4}HKUST(GZ)\\
    \textsuperscript{*}Corresponding Author:
    \href{mailto:wei.zhang@ust.hk}{wei.zhang@ust.hk}
  }
  \country{}
}

\renewcommand{\shortauthors}{}



\begin{abstract}

3D reconstruction and view synthesis are fundamental to AR/VR, robotics, and digital twins. The Visual Geometry Grounded Transformer (VGGT) enables strong feed-forward 3D reconstruction while its billion-parameter scale limits on-device deployment. LLM-oriented quantization methods fail on VGGT due to saturated activation channels that resist low-bit quantization and diverse 3D semantics that impede calibration. VGGT further poses hardware challenges from multi-precision architecture support and long-sequence global attention with excessive memory demands.
We propose VersaQ-3D, an algorithm-architecture co-design framework for efficient VGGT inference. \uline{At the algorithm level}, we present the first calibration-free, input-agnostic quantization method for VGGT, leveraging transform coding to suppress outliers and preserve structural weight features, enabling robust low-bit inference down to 4 bits. \uline{At the architecture level}, we design a reconfigurable accelerator with a hierarchical multi-precision compute unit (BF16/INT8/INT4) that executes both linear and non-linear operators within a shared systolic datapath, reducing end-to-end latency by 77\%. A two-stage recomputation-based tiling strategy further cuts runtime by 7\% by alleviating on-chip memory pressure for long-sequence attention.
Evaluations across various datasets show that VersaQ-3D incurs negligible accuracy loss at W4A8 and consistently achieves leading accuracy at W4A4 over prior quantization methods across diverse scenes. The co-designed accelerator delivers 5.4$\times$-22.0$\times$ speedup over edge GPUs and 2.2$\times$-3.0$\times$ over prior quantization-based accelerators under iso-PE-area comparison, enabling instant and energy-efficient feed-forward 3D reconstruction on edge devices.

\end{abstract}



\maketitle

\section{Introduction}\label{sec:intro}

3D reconstruction and view synthesis are fundamental to computer vision and graphics, with applications spanning AR/VR~\cite{zhao2020deja,li2023instant}, robotics and autonomous systems~\cite{zhou2017unsupervised,zhu2022nice}, digital twins~\cite{mihai2022digital}, and smart city modeling~\cite{deng2021systematic}. As these applications scale to larger and more dynamic environments, reconstruction methods must be not only accurate but also computationally efficient and robust across diverse scenarios.

Traditional methods such as Structure from Motion (SfM)~\cite{schonberger2016structure} and Multi-View Stereo (MVS)~\cite{hartley2003multiple} rely on iterative, multi-stage optimization.
These methods are well established and underpin many real-world systems, including mapping and localization in autonomous driving, but their optimization can be computationally expensive and slow to converge, which poses challenges for latency-sensitive or resource-constrained deployment~\cite{zhang2025advances}.
Learning-based representations such as Neural Radiance Fields (NeRF)~\cite{mildenhall2021nerf} and 3D Gaussian Splatting (3DGS)~\cite{kerbl20233d} improve rendering quality but still require per-scene optimization and focus primarily on view synthesis rather than explicit geometric reconstruction, limiting their use in downstream perception and robotics tasks. In contrast, feed-forward approaches directly produce explicit 3D outputs, including camera poses, depth maps, and point maps, from input images in a single forward pass without iterative refinement. The Visual Geometry Grounded Transformer (VGGT)~\cite{wang2025vggt} exemplifies this paradigm: it reconstructs 3D geometry from one to dozens of images in tens of seconds, often matching or surpassing traditional methods without any post-processing.

\begin{figure}[!t]
     \centering
    
    \includegraphics[width=\linewidth]{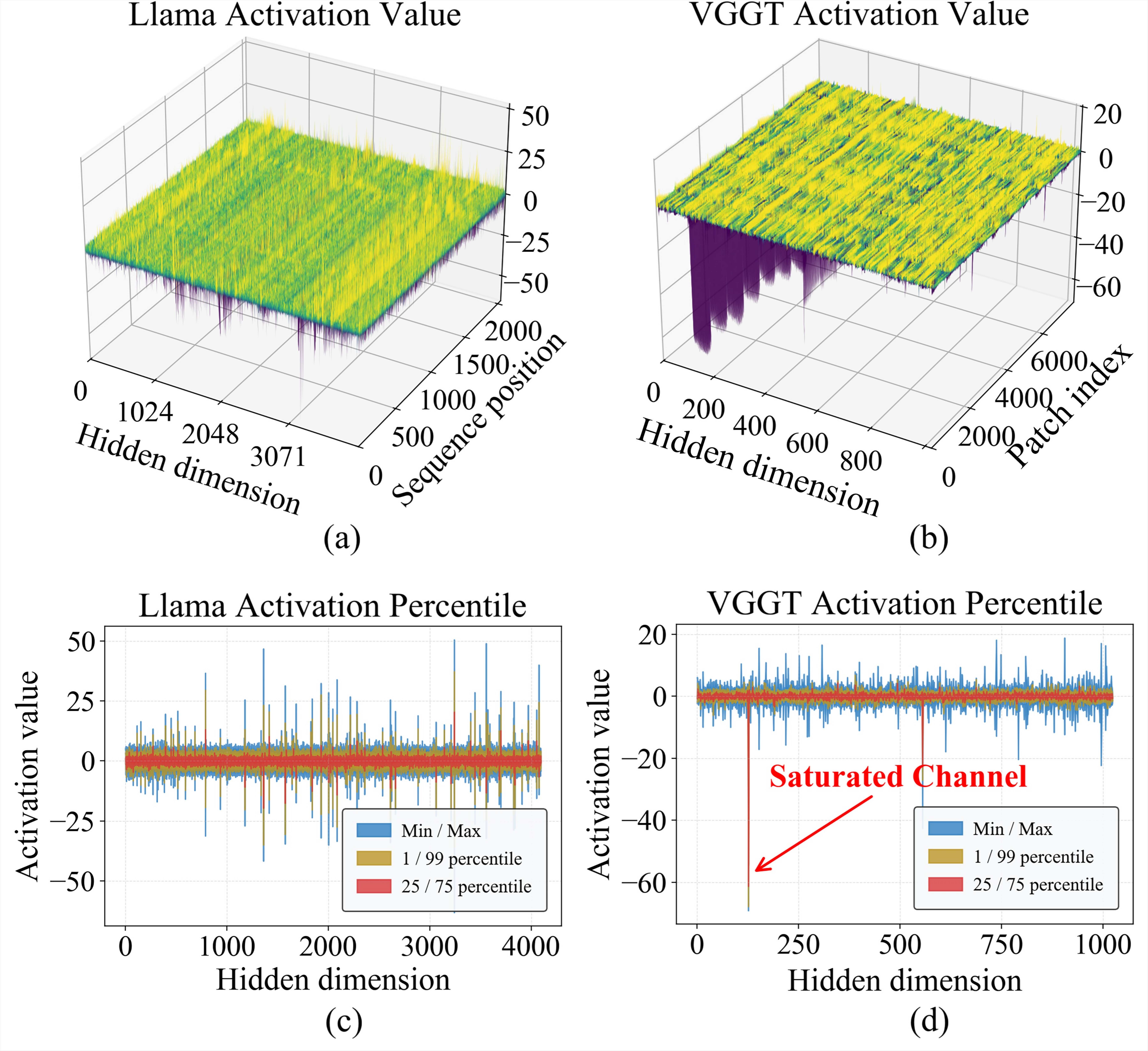}
    
    \caption{Llama and VGGT activation value distributions and corresponding percentile distributions. The red band (25th-75th percentiles) denotes the inter-quartile range containing the middle 50\% of activations. Unlike LLMs, which show isolated spiking outliers, VGGT exhibits saturated activation channels where many channels maintain persistently high activation values.}
    \label{fig:activation}
    
\end{figure}

Despite its remarkable reconstruction capability, the billion‑scale parameters of VGGT pose substantial challenges to achieving instant 3D reconstruction~\cite{li2023instant,muller2022instant}, where high‑fidelity results are expected within only a few seconds of visual input~\cite{miller1968response,nah2004study,bonci2021human}. However, current transformer‑based 3D models fall far short of such latency targets due to their extreme memory and compute demands. VGGT consists of over 1.2B parameters, requiring $>$ 4 GB for its FP32 weights and more than 7 TFLOPs of computation per inference for multi‑view reconstruction. A typical edge GPU such as the NVIDIA Jetson Orin NX~\cite{ditty2022nvidia}, which offers 19.2 FP16 TFLOPS of peak compute but only limited LPDDR5 memory with 102.4 GB/s bandwidth, meaning that raw compute throughput alone does not determine end-to-end latency. Profiling results (Sec.~\ref{subsec:challenge}) reveal that VGGT is constrained primarily by memory capacity and bandwidth, while the quadratic cost of global attention further amplifies latency as the number of input views increases. Moreover, edge GPUs such as the Jetson Orin NX operate at a typical power consumption of 25 W, whereas the power constraints necessary for devices like AR glasses are under 10 W~\cite{noh2025flexnerfer}. Consequently, the combination of massive parameter size, bandwidth sensitivity, quadratic computation cost, and excessive power consumption of existing edge GPUs makes direct on‑edge deployment of VGGT impractical without model compression and hardware support.

Quantization, which reduces model size and computational cost by lowering numerical precision, appears to be a natural path to improve VGGT's efficiency. Since VGGT shares the transformer backbone with modern large language models (LLMs), many existing LLM‑oriented quantization techniques appear applicable. However, these methods are ineffective for VGGT due to two fundamental differences from LLMs in both model behavior and data characteristics: \textbf{First}, VGGT exhibits \textbf{\textit{saturated activation channels}}, where many channels maintain persistently high activation values rather than isolated spiking outliers. This saturation behavior, illustrated in Fig.~\ref{fig:activation}, contrasts sharply with the isolated spiking outliers observed in LLMs and makes existing outlier‑smoothing techniques~\cite{ashkboos2024quarot,liu2024spinquant} ineffective at maintaining model accuracy after quantization. \textbf{Second}, the \textbf{\textit{diverse and scene‑dependent semantics of 3D multi‑view data}} hinder effective calibration~\cite{feng2025quantized}: small datasets overfit to specific inputs, while large ones remain costly and fail to cover unseen geometries. These challenges prevent standard post‑training quantization pipelines from maintaining VGGT’s reconstruction accuracy or generalizing well across dynamic 3D environments.

In addition, VGGT's multi-task nature makes precision support more challenging. Camera pose and depth prediction can tolerate aggressive quantization, while point map estimation is much more sensitive to low precision~\cite{zhang2026not}. The non-linear operators in the model, such as LayerNorm, Softmax, and GELU, are also highly sensitive to numerical precision and often require BF16-level computation to preserve geometric accuracy. This creates a \textbf{\textit{mixed-precision dilemma}}: low-bit linear layers are needed for efficiency~\cite{yu20248}, while high-precision non-linear operators and precision-sensitive tasks are needed for accuracy. Supporting these precision modes with separate hardware units introduces \textbf{\textit{additional hardware resource overhead}} and may cause workload imbalance, while a fixed-precision design cannot fully exploit the efficiency of quantized computation.
Furthermore, the global attention mechanism in VGGT introduces another challenge. Unlike autoregressive LLM decoding, VGGT performs a single forward pass that processes all input frames simultaneously by concatenating them into a \textit{\textbf{single long sequence}}. The full quadratic attention cost is therefore incurred at once, producing extremely large activation tensors and intensive matrix multiplications within a single pass. This places heavy demands on both memory bandwidth and computation units, and as we discuss in Section~\ref{subsec:tiling}, standard tiling strategies such as FlashAttention~\cite{dao2024flashattention} face unique difficulties under these conditions.

To address the aforementioned inefficiencies, we propose VersaQ-3D, an algorithm-architecture co-design framework for efficient feed-forward 3D reconstruction. Our main contributions are summarized as follows:

\begin{itemize}

\item We systematically profile and analyze the deployment bottlenecks of VGGT on edge devices. We are the first to discover that saturated activation channels and diverse 3D task semantics are the key obstacles to effective quantization, while multi-precision requirements and long input sequences pose critical challenges for hardware design.

\item Guided by these findings, we propose VersaQ-3D, the first calibration-free, input-agnostic post-training quantization framework for VGGT. It employs integer-friendly transform coding to mitigate activation saturation and preserve structural weight features, enabling effective quantization down to 4 bits where existing PTQ methods struggle to maintain accuracy.

\item On the architecture side, we design a reconfigurable accelerator featuring a hierarchical multi-precision compute unit (BF16/INT8/INT4) that executes linear, non-linear, and transform coding operators within a shared systolic datapath. A two-stage recomputation-based tiling scheme further alleviates the long-sequence memory bottleneck.

\item Experiments across diverse datasets (Co3Dv2~\cite{reizenstein2021common}, RealEstate10K~\cite{zhou2018stereo}, 7-Scenes~\cite{shotton2013scene}, and DTU~\cite{jensen2014large}) show that VersaQ-3D incurs negligible accuracy loss compared to full precision at W4A8 and consistently achieves state-of-the-art accuracy at W4A4 over prior quantization methods. The co-designed accelerator delivers 5.4$\times$-22.0$\times$ speedup over edge GPUs and 2.2$\times$-3.0$\times$ speedup over prior quantization-based accelerators (ANT~\cite{guo2022ant}, Olive~\cite{guo2023olive}) under iso-PE-area comparison. For energy efficiency, the accelerator achieves 63.1$\times$-202.0$\times$ gains over edge GPUs (Jetson Xavier NX and Orin NX) and 2.5$\times$-3.5$\times$ gains over prior quantization-based accelerators, enabling instant and energy-efficient 3D reconstruction on edge devices.

\end{itemize}

\section{Background and Motivation}\label{sec:background}
\subsection{Visual Geometry Grounded Transformer}\label{sec:vggt}

\begin{figure}[!t]
    \centering
    \includegraphics[width=.9\linewidth]{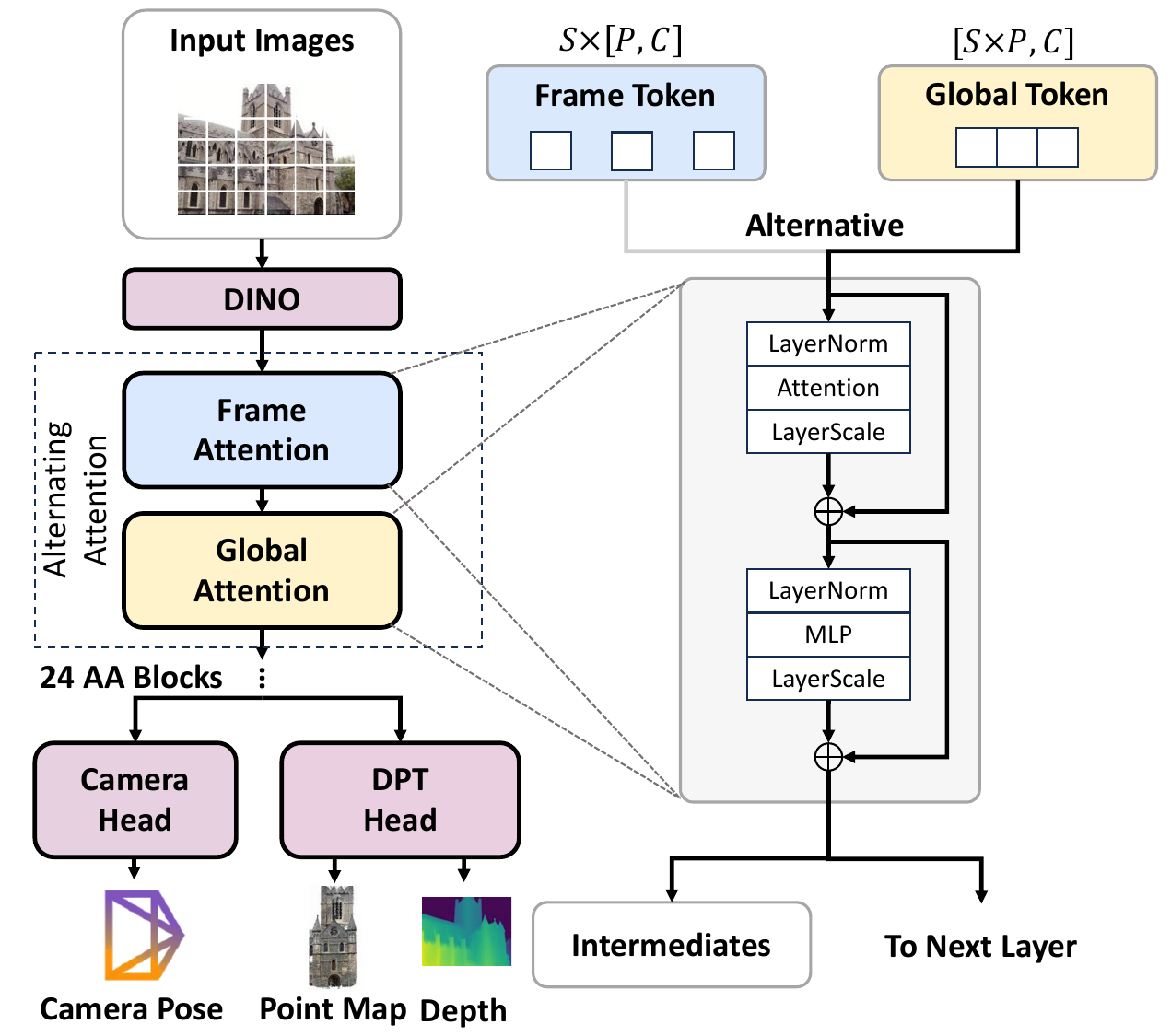}
    \caption{VGGT model structure. A DINO-based feature extractor feeds an AA module that interleaves frame and global attention over tokens reshaped between $S\times[P,C]$ and $[S\times P, C]$.}
    \label{fig:vggt}
\end{figure}

VGGT is composed of several key components arranged in a structured pipeline as shown in Fig.~\ref{fig:vggt}. It begins with DINO~\cite{oquab2024dinov2} for feature extraction, followed by an Alternating-Attention (AA) transformer. The AA module alternates between two types of self-attention: frame attention for intra-frame analysis and global attention for inter-frame integration. This is achieved by manipulating the token tensor, which has a shape of $[B, S, P, C]$ (Batch, Frames, Patches, Channels). For global attention, the tensor is reshaped to $[B, S\times P, C]$, treating all patches from all frames as a single sequence. For frame attention, it is processed as $S$ parallel sequences of shape $[B, P, C]$, confining attention within each frame. On top of the attention backbone, VGGT attaches task‑specific prediction heads. For 3D reconstruction, this typically includes a Camera Head for estimating camera poses and a DPT Head for producing depth maps and reconstructing point maps. 

\subsection{Transform Coding}
Transform coding is fundamental for signal decorrelation and energy compaction, providing a mathematical basis for efficient computation in both classic and modern architectures~\cite{ahmed2012orthogonal}. For any orthogonal transform represented by matrix $\mathbf{Q}$, the relation $\mathbf{Q}\mathbf{Q}^T = \mathbf{I}$ holds. Here, we introduce two representatives that we adopt in our work: WHT~\cite{pratt1969hadamard} and DCT~\cite{khayam2003discrete}. The WHT is a linear, orthogonal, and symmetric transform represented by a Hadamard matrix $\mathbf{H}$, satisfying $\mathbf{H} = \mathbf{H}^T$ and $\mathbf{H}\mathbf{H}^T = \mathbf{I}$ for its normalized form. The matrix can be recursively constructed as
\begin{equation}
\mathbf{H_2} = \frac{1}{\sqrt{2}}
\begin{bmatrix}
1 & 1 \\
1 & -1
\end{bmatrix},
\quad
\text{where } \mathbf{H_{2^n}} = \mathbf{H_2} \otimes \mathbf{H_{2^{(n-1)}}}.
\end{equation}
Therefore, its entries are fixed to $\pm1$.
In contrast, the DCT projects input data onto a set of cosine basis functions, achieving strong energy compaction that has made it a core component of compression standards~\cite{ITUH265}. In our design, the DCT captures the structural features of weights in a compact frequency-domain representation, so that quantization in this domain better preserves these structures after inverse transformation. Because both bases are fixed rather than learned from data, they may be less effective for decorrelation on a specific input. In return, they require no training and no extra storage for the basis, and they support fast transforms, which makes them well-suited to our setting.

\subsection{Observations and Challenges}\label{subsec:challenge}

\begin{figure}
    \centering
    
    \begin{minipage}[b]{0.23\textwidth}
        \centering
        \subfloat[Different GPUs (S=3)\label{fig:profile_gpu}]{
            \includegraphics[width=.9\textwidth]{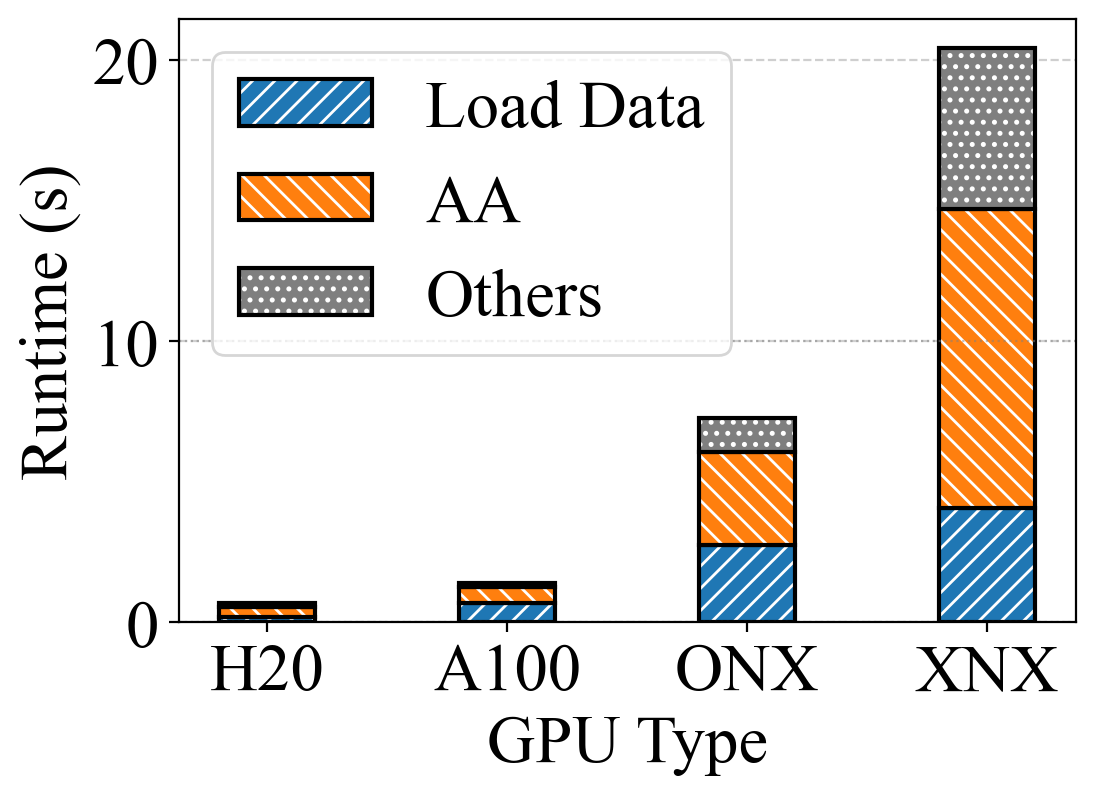}
        }
    \end{minipage}
    \hfill
    \begin{minipage}[b]{0.23\textwidth}
        \centering
        \subfloat[Different S on Jetson Orin NX\label{fig:profile_s}]{
            \includegraphics[width=.9\textwidth]{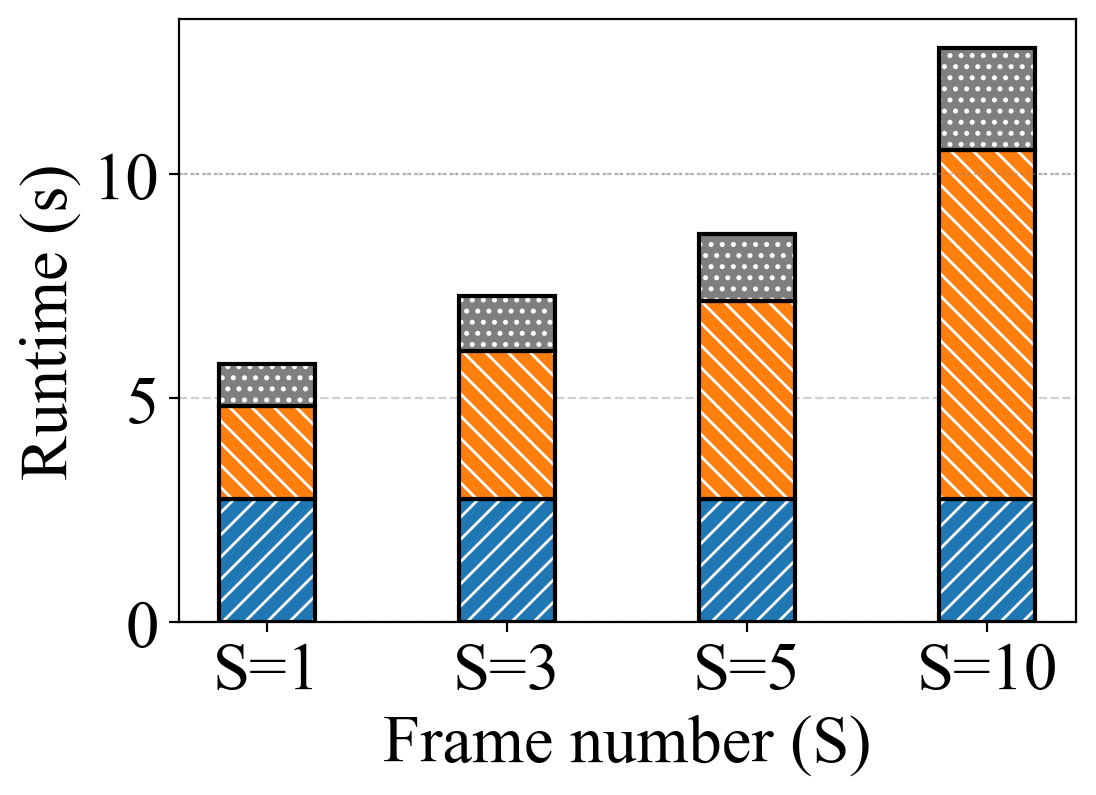}
        }
    \end{minipage}

    \caption{Inference runtime breakdown on (a) different GPUs with S=3 and (b) different sequence lengths S on Jetson Orin NX, evaluated on the Example/Kitchen dataset.}
    \label{fig:profile}
\end{figure}

\textbf{Observations.} To motivate the co-design of quantization and hardware acceleration, we first profile VGGT's runtime behavior across diverse GPU platforms. As shown in Fig.~\ref{fig:profile}, the runtime of VGGT can be divided into
three major components: model weight loading, the AA module, and other
operations. On edge GPUs such as Jetson Orin NX 16\,GB
(ONX)~\cite{ditty2022nvidia} and Jetson Xavier NX 16\,GB
(XNX)~\cite{ditty2018nvidia}, the weight-loading stage dominates total
latency, whereas on server GPUs (H20 96\,GB~\cite{zhu2025megascale} and A100
40\,GB~\cite{choquette2020nvidia}) the high-bandwidth memory mitigates this
overhead. Unlike LLMs, where parameter loading is amortized over long
sequential decoding~\cite{zhang2024llmcompass,heo2024neupims}, VGGT performs a single forward pass that processes all
input frames simultaneously, making the loading latency a dominant contributor to end-to-end
runtime on LPDDR-based edge
systems~\cite{JEDEC_JESD209_4,JEDEC_JESD209_5C} whose bandwidth is nearly
two orders of magnitude lower than HBM2~\cite{JEDEC_JESD235D,larimi2021understanding} or
HBM3~\cite{JEDEC_JESD238B01,son2023thermal} counterparts
(Fig.~\ref{fig:profile}(a)). Meanwhile, the global attention in the AA
module introduces quadratic growth in computation and memory usage with
respect to the number of frames~$S$, since the attention score matrix scales
as $[S \times P,\; S \times P]$, resulting in rapidly increasing latency as
sequence length grows (Fig.~\ref{fig:profile}(b)). 
To address the weight-loading bottleneck, an effective approach is aggressive model quantization to reduce the memory footprint. However, as we will show, VGGT's distinct activation patterns and diverse 3D task semantics make low-bit quantization non-trivial. On the hardware side, the quadratic attention cost and the mixed-precision demands of non-linear operators require careful architectural design. We discuss these two sets of challenges below.

\begin{figure}[!t]
    \centering
    
    \includegraphics[width=.5\textwidth]{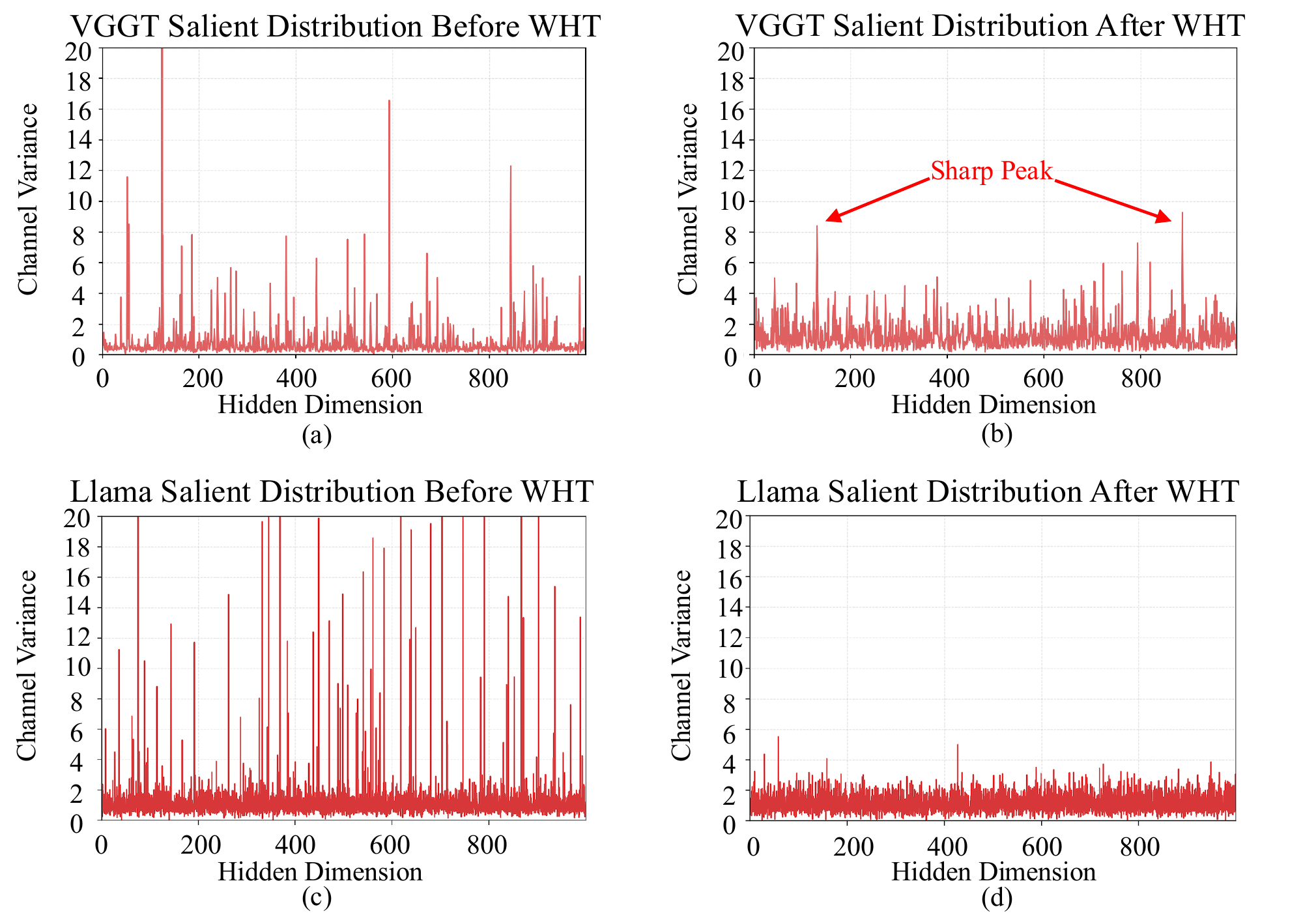}

    \caption{VGGT and Llama salient distribution comparison before and after WHT. Here we take channel variance to indicate 
    the saliency of the activation. VGGT's variance exhibits considerable peaks after 
    Hadamard rotation.}
    \label{fig:salient}
\end{figure}

\textbf{Quantization Challenges.}
\textit{\textbf{Challenge~I: Saturated Activation Channels Limit Low-Bit
Quantization.}}
As shown in Fig.~\ref{fig:activation}, VGGT exhibits highly saturated
activation channels, where many channels maintain uniformly large activation
magnitudes across a wide percentile range. In contrast, LLMs such as
LLaMA~\cite{touvron2023llama} display isolated spiking outliers, where
extreme activations mainly occur near the min-max range while the
25th--75th percentiles remain moderate. This saturation pattern indicates
that large activation values persist across most percentiles within certain
channels rather than appearing as isolated extremes. Consequently, prior
rotation-based quantization techniques~\cite{ashkboos2024quarot,liu2024spinquant},
which leverage transforms such as the WHT to redistribute and smooth isolated
outliers, become less effective for VGGT. As illustrated in
Fig.~\ref{fig:salient}, after Hadamard rotation the overall channel-wise
variance profile becomes milder, yet noticeable peaks remain across multiple
channels, revealing that considerable activation variation still persists.
This distinctive behavior undermines existing quantization methods designed
for LLMs and necessitates an approach that not only redistributes activation
outliers but also preserves the structural weight features associated with
these high-variance channels.

\textit{\textbf{Challenge~II: Diverse 3D Semantics Impede
Calibration-Based Quantization.}}
Calibration-dependent quantization
methods~\cite{lin2024awq,xiao2023smoothquant,dettmers2022gpt3} rely on a
small, representative dataset to statistically profile activation and weight
distributions, identifying data characteristics and other parameters
essential for determining quantization configurations. However, different 3D
scenes occupy highly distinct regions of the model's feature
space~\cite{feng2025quantized}, making it infeasible for any small
calibration set to capture the full data distribution. While enlarging the
calibration set may seem like a solution, it introduces prohibitive
computational overhead for pre-inference calibration and still offers no
guarantee of covering the feature distributions of novel, unseen geometries.
Consequently, calibration-based approaches lead to unstable parameter
estimation and overfitting to the calibration data, ultimately causing
significant degradation in reconstruction fidelity on unseen scenes. This
challenge calls for a calibration-free quantization strategy that is
inherently input-agnostic.

\textbf{Hardware Challenges.}
\textit{\textbf{Challenge~III: Mixed-Precision Requirements Cause Hardware
Underutilization.}}
The non-linear operators in VGGT, including LayerNorm, Softmax, and GELU,
exhibit high sensitivity to numerical precision and require BF16-level
computation to maintain accuracy. Meanwhile, the dominant linear operators
(\textit{e.g.}, matrix multiplications in projections and attention) benefit
significantly from aggressive low-bit INT execution. This mixed-precision
requirement poses a hardware design dilemma: dedicating separate
floating-point units to non-linear operators introduces additional area and
power overhead, while a fixed-precision architecture cannot simultaneously
exploit the efficiency of low-bit computation for linear layers and the
accuracy of higher precision for non-linear layers. A reconfigurable compute 
unit that supports multiple precisions within the
same datapath is therefore essential.
\textit{\textbf{Challenge~IV: Long-sequence global attention incurs
excessive off-chip traffic.}}
The Softmax operation in the AA module requires row-wise normalization
over all key positions, creating a global dependency across \(K,V\) tiles.
FlashAttention~\cite{dao2024flashattention} reduces output traffic on GPUs by keeping each output
tile \(O_i\) in on-chip SRAM, updating it across all \(K,V\) tiles, and
writing it to HBM only once. This assumption is hard to satisfy on
systolic-array accelerators~\cite{lin2025systolicattention}, where limited on-chip buffers are also needed for data movement and intermediate results of \(QK^{T}\) and \(SV\), causing
\(O_i\) to be unable to stay resident during the whole attention computation. As a
result, each \(K,V\) tile requires reading, updating, and writing back
\(O_i\), causing \(O(N^2)\) off-chip traffic. This traffic is especially
costly on edge devices with bandwidth-limited LPDDR~\cite{JEDEC_JESD209_5C, horowitz20141}. A memory-efficient tiling strategy that avoids iterative off-chip updates of \(O_i\) is
therefore essential.

\section{Versatile Transform Coding Based Quantization}

This section presents our versatile transform-coding-based quantization framework for VGGT. We first outline the design considerations motivated by the quantization challenges identified in Sec.~\ref{subsec:challenge}, explaining how transform coding addresses both saturated activations and calibration dependency while remaining hardware-friendly (Sec.~\ref{subsec:dc}). We then detail the two-phase quantization pipeline, consisting of offline weight preparation and online inference, that integrates WHT and DCT to produce quantization-friendly representations with minimal runtime cost (Sec.~\ref{subsec: quantization}).

\subsection{Design Considerations}\label{subsec:dc}

Our quantization framework is guided by three principles derived from the challenges in Sec.~\ref{subsec:challenge}. First, to address \textit{\textbf{saturated activation channels (Challenge I)}}, we adopt a dual-transform strategy: an orthogonal WHT redistributes saturated outliers across feature dimensions following the incoherence principle~\cite{tseng2024quip}, while a DCT on weights compacts essential structural features, particularly the components associated with high-variance channels, into a small number of low-frequency coefficients, mitigating the residual quantization error that redistribution alone leaves unresolved. Second, to eliminate \textit{\textbf{calibration dependency (Challenge II)}}, both transforms are fixed and data-independent, making the framework inherently input-agnostic regardless of the diversity of 3D geometries encountered at inference time. Third, to maintain \textit{\textbf{hardware efficiency}}, we select integer-friendly transform implementations. Specifically, WHT entries are constrained to $\pm1$ (requiring only additions), and the DCT adopts the HEVC/H.265 integer matrix~\cite{ITUH265}. Moreover, we fuse both transforms into the offline weight preparation stage (Sec.~\ref{subsec: quantization}), adding negligible runtime overhead.

\subsection{Quantization Framework} \label{subsec: quantization}
\begin{figure*}
    \centering
    \includegraphics[width=\linewidth]{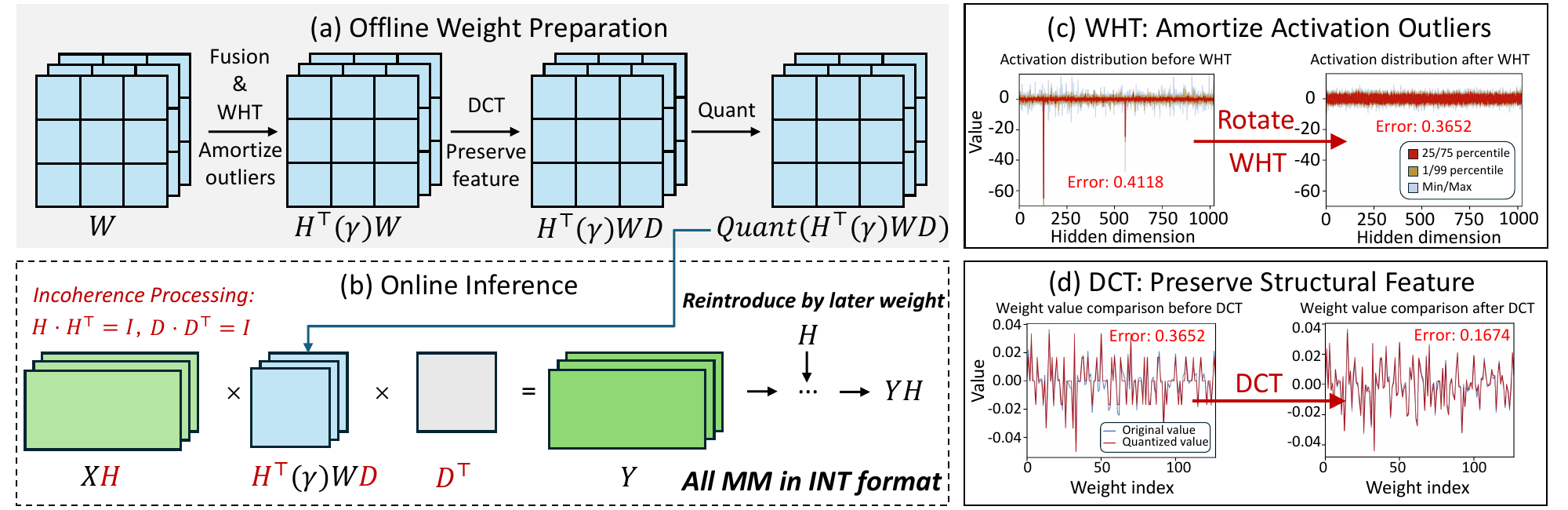}
    \caption{VersaQ-3D algorithm overview. VersaQ-3D pairs two data-independent transform coding methods, WHT for activations and DCT for weights, enabling calibration-free, input-agnostic quantization. (a) Offline weight preparation fuses the WHT, scaling factors, and DCT into the weight matrix, incurring zero runtime cost. (b) Online inference cancels the paired transforms through the linear projection via the incoherence processing principle. (c) WHT redistributes saturated activation channels into a uniform distribution via incoherence. (d) DCT transforms weights into the frequency domain, where quantization better preserves structural weight features compared to direct spatial-domain quantization.}
    \label{fig:algorithm}
\end{figure*}

VersaQ-3D integrates WHT and DCT through a two-phase pipeline---offline weight preparation and online inference---as illustrated in Fig. ~\ref{fig:algorithm}(a) and (b). The framework applies to all linear projections in both the attention module (Q, K, V, and output projections) and the FFN, exploiting orthogonal transform properties to produce quantization-friendly representations while preserving mathematical equivalence.

\textbf{Mathematical Foundation.} For any linear layer computing $Y=XW$, inserting matched orthogonal pairs yields:
\begin{equation}
    (X\mathbf{H})(\mathbf{H}^{\top}W\mathbf{D})(\mathbf{D}^{\top}) = X(\mathbf{HH}^{\top})W(\mathbf{D}\mathbf{D^{\top}}) = XW.
\end{equation}
This identity ensures that transforming activations by $\mathbf{H}$ and weights by $\mathbf{H^{\top}}$ and $\mathbf{D}$ does not alter the computation, provided the inverse transform $\mathbf{D^{\top}}$ is applied to recover the output. Moreover, this equivalence extends to residual connections: because both the residual and the sub-layer output are maintained in the rotated domain throughout the network, their element-wise addition remains valid without additional transforms, i.e. $X\mathbf{H} + Y\mathbf{H} = (X + Y)\mathbf{H}$.

\textbf{Offline Weight Preparation.} As shown in Fig.~\ref{fig:algorithm}(a), each weight matrix is processed in three steps before deployment. First, $\mathbf{H^{\top}}$ is applied along the input dimension to pair with the Hadamard-rotated activation that will arrive during inference. For layers that include element-wise scaling, such as the scaling factor $\gamma$ in LayerNorm and LayerScale, these parameters are fused into the weight at this stage, yielding $\mathbf{H}^{\top}(\gamma)W$. This fusion is valid because the orthogonal pair $\mathbf{H}/\mathbf{H}^{\top}$ cancels during the subsequent linear projection, so the scaling factor can be equivalently applied on the weight side without altering the output. Second, the DCT matrix $\mathbf{D}$ is applied along the output dimension, projecting the weight into the frequency domain. As illustrated in Fig.~\ref{fig:algorithm}(d), quantization in the
frequency domain better preserves the structural features of the original weights compared to direct spatial-domain quantization. Third, the fully transformed weight $\mathbf{H}^{\top}(\gamma)W\mathbf{D}$ is quantized to low-bit INT format for storage and deployment. Since all three steps are performed once offline, they incur zero runtime cost.


\textbf{Online Inference.} During inference (Fig.~\ref{fig:algorithm}(b)), the input activation $X\mathbf{H}$—pre-rotated from the preceding layer—is multiplied with the quantized weight in low-precision INT arithmetic. An IDCT ($\mathbf{D}^{\top}$) is then applied to recover the output from the frequency domain, yielding $Y=XW$. The output is subsequently rotated by $\mathbf{H}$ to produce $Y\mathbf{H}$, which serves directly as the pre-rotated input for the next layer. This cross-layer rotation chaining eliminates redundant WHT operations at every layer boundary: only the IDCT and a single WHT are required per layer, minimizing online computation overhead. As shown in Fig.~\ref{fig:algorithm}(c), the WHT effectively redistributes the saturated activation channels identified in Sec.~\ref{subsec:challenge} into a more uniform distribution, enabling accurate low-bit activation quantization without calibration data. For attention-specific operations that require activations in the original (unrotated) domain, such as RoPE, the IDCT first recovers the spatial-domain representation, the operator is applied in BF16 precision, and the WHT is re-applied before re-quantization to INT. Similarly, non-linear operations including LayerNorm, GELU, and Softmax are also performed in BF16 precision, requiring additional precision transitions at their boundaries. These transitions are limited to a small number of operators per layer and do not dominate overall runtime.

\textbf{Quantization Configuration.}
VersaQ-3D adopts symmetric uniform quantization at token-wise
granularity for activations and channel-wise granularity for weights.
As a post-training quantization (PTQ) method, it requires no
training or fine-tuning.
We consider two bit-width configurations:
W4A8 as the \emph{conservative} setting that prioritizes reconstruction
fidelity, and W4A4 as the \emph{aggressive} setting that prioritizes
hardware efficiency.
Both configurations share the same offline weight preparation
and online inference pipeline described above.

\section{Reconfigurable Accelerator for VersaQ-3D}

This section presents the reconfigurable accelerator architecture of VersaQ-3D. We first outline the design considerations motivated by the hardware challenges identified in Sec.~\ref{subsec:challenge}, explaining how hierarchical composition and recomputation-based tiling jointly address the mixed-precision dilemma and long-sequence memory issue (Sec.~\ref{subsec:hw_dc}). We then detail the overall architecture and its hierarchical multi-precision PE design, which supports INT4, INT8, and BF16 execution within a single reconfigurable datapath (Sec.~\ref{subsec:overall} and Sec.~\ref{subsec:pe}). Finally, we present the pipelined quantization and dequantization units for precision transitions (Sec.~\ref{subsec:qdq}) and the two-stage tiling strategy that reduces off-chip traffic for the output matrix from quadratic to linear in the sequence length (Sec.~\ref{subsec:tiling}).

\begin{figure}[t]
    \centering
    \includegraphics[width=\linewidth]{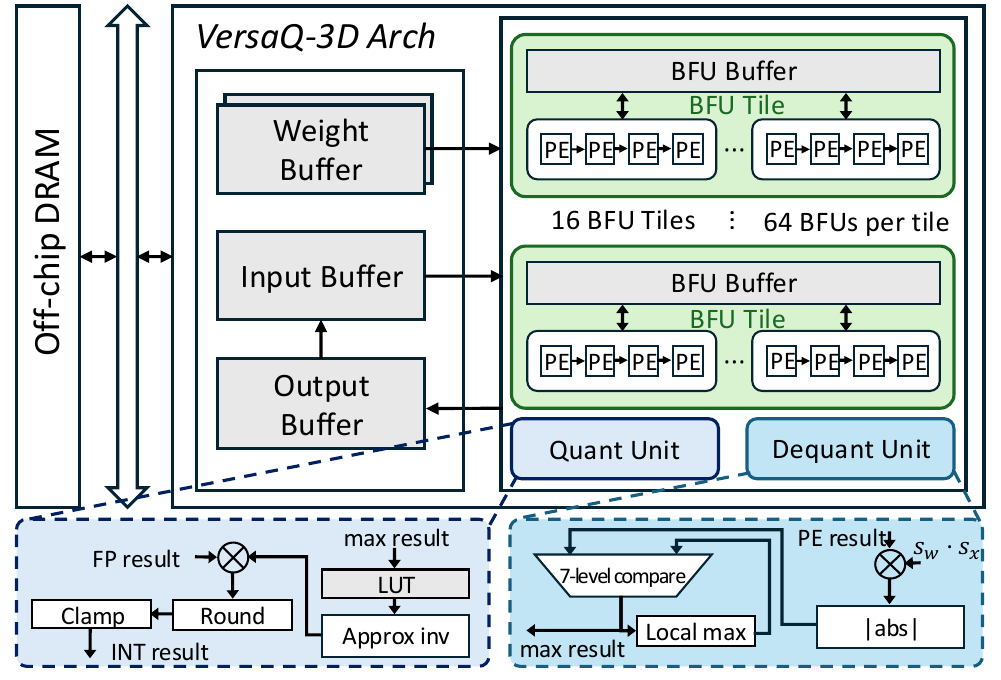}
    \caption{Overall architecture of VersaQ-3D accelerator. The lower portion details the online quantization and dequantization units, enabling on-the-fly requantization without off-chip memory round trips.}
    \label{fig:arch}
\end{figure}

\subsection{Design Considerations}\label{subsec:hw_dc}

The accelerator design is guided by two principles derived from the hardware challenges identified in Sec.~\ref{subsec:challenge}, mirroring the structure of our algorithm design considerations in Sec.~\ref{subsec:dc}. First, to resolve the \textit{\textbf{mixed-precision dilemma (Challenge III)}}, we pursue a hierarchical composition strategy that supports INT4, INT8, and BF16 execution within a single reconfigurable datapath. INT4 PEs serve as atomic building blocks: groups of four compose one INT8 PE through bit-fusion, and four INT8 PEs with lightweight peripheral logic form one Bitwidth Flexible Unit (BFU) capable of BF16 arithmetic. This three-level hierarchy allows the same physical hardware to dynamically switch among precision modes without dedicated floating-point units. Each INT4 PE further integrates a dual-mode multiplexer co-designed with the online WHT (Sec.~\ref{subsec: quantization}), folding the Hadamard transform directly into the datapath. Second, to mitigate the excessive on-chip memory demand of \textit{\textbf{long-sequence global attention (Challenge IV)}}, we adopt a two-stage recomputation-based tiling strategy that decouples Softmax statistic accumulation from output computation. The first stage scans all
K tiles to finalize the row-wise Softmax statistics with only compact intermediate vectors kept on-chip. The second stage recomputes the scores, applies Softmax with the finalized statistics, and writes each completed output tile to off-chip memory exactly once. This trades inexpensive INT-precision recomputation for a constant peak memory footprint independent of sequence length.

\begin{figure*}
    \centering
    \includegraphics[width=\linewidth]{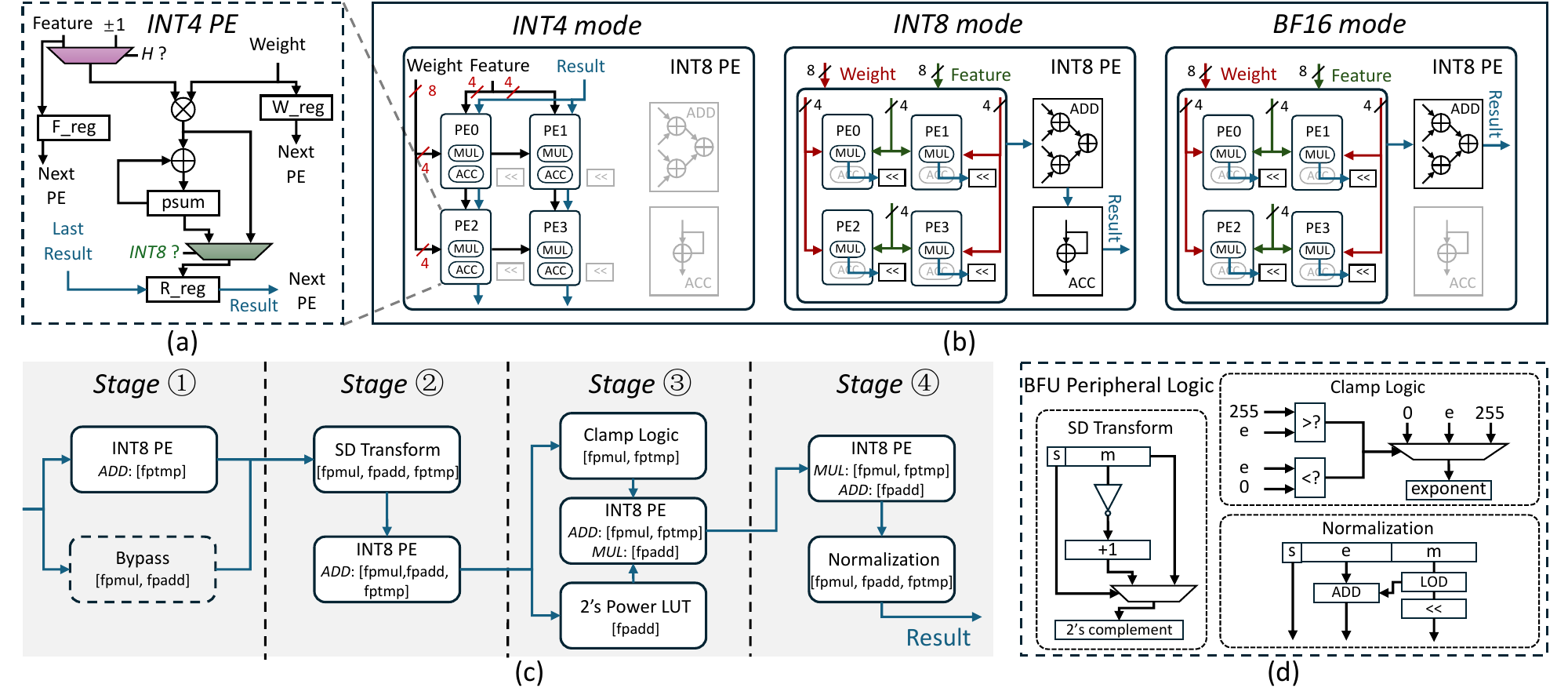}
    \caption{Unified PE architecture that composes INT4 PEs to support INT4, INT8, and BF16 computation without dedicated floating-point hardware. (a) INT4 PE as the atomic building block. (b) INT8 PE operating in three precision modes with selective datapath activation. (c) BF16 execution decomposed into a 4-stage pipeline that reuses INT8 PEs for mantissa/exponent arithmetic, with SD Transform, clamp logic, and normalization handling format conversion. (d) Peripheral logic details for the BF16 stages. This reconfigurable design serves all precisions needed in our quantized model, minimizing area overhead.}
    \label{fig:pe_design}
\end{figure*}

\subsection{Overall Architecture}\label{subsec:overall}

Fig.~\ref{fig:arch} presents the overall architecture of VersaQ-3D. 
The compute core is organized as 16 BFU Tiles, each containing 64 BFUs. Each BFU comprises 4 INT8 PEs, and each INT8 PE in turn contains 4 INT4 PEs, yielding an equivalent capacity of 64$\times$64 INT8 PEs or 128$\times$128 INT4 PEs depending on the active precision mode. The on-chip memory system includes a double-buffered weight buffer, single-buffered input and output buffers, and one dedicated BFU buffer per tile for storing BFU inputs and outputs. In INT4 and INT8 modes, the array operates as an output-stationary systolic array where partial sums accumulate locally within each PE, minimizing data movement for the dense matrix multiplications in attention projections and feed-forward layers. In BF16 mode, the architecture reconfigures into a SIMD-style vector unit: each BFU fetches operands from its tile's BFU buffer, and results are either written back to the BFU buffer for multi-stage operations or forwarded to the output buffer. The Quantization and Dequantization units in the lower portion of Fig.~\ref{fig:arch} manage precision transitions and are detailed in Sec.~\ref{subsec:qdq}.

\subsection{Hierarchical Multi-Precision PE}\label{subsec:pe}

Fig.~\ref{fig:pe_design} illustrates the hierarchical PE design, which spans three levels: INT4 PE, INT8 PE, and BFU.

\textbf{INT4 PE.} The INT4 PE is the atomic compute unit, performing a single 4-bit MAC per cycle using a 4-bit multiplier and an 8-bit adder. The result is saved in the 8-bit \textit{R\_reg} register. As shown in Fig.~\ref{fig:pe_design}(a), it features a dual-mode input multiplexer. In matrix-multiplication mode, the feature input arrives from the systolic interconnect. In WHT mode, the multiplexer directly configures $\pm 1$ Hadamard coefficients, bypassing the weight path. This avoids both the SRAM overhead of storing the Hadamard matrix and the bandwidth contention of fetching it through the shared weight path.

\textbf{INT8 PE.} Fig.~\ref{fig:pe_design}(b) shows the INT8 PE operating in its three precision modes with the corresponding dataflow and active hardware units highlighted. Four INT4 PEs compose one INT8 PE and reuse their INT4 multiplier. In INT4 mode, the four INT4 PEs function as a $2\times 2$ output-stationary systolic array, each operating independently on a separate operand pair. In INT8 mode, the input feature and weight are each split into high and low 4-bit segments, distributed to the four INT4 PEs for partial product computation, and then combined through shifting and an adder tree to produce the full 8-bit product. In BF16 mode, the datapath computes integer products from decomposed BF16 mantissa and exponent components identically to INT8 mode, except that the local accumulator is bypassed since the BFU's peripheral logic handles result composition externally. This design allows the INT8 PE to serve as both a standalone integer MAC unit in systolic mode and a building block for BF16 execution in SIMD mode, enabling the entire array to transition between precision modes without dedicating hardware units exclusively to floating-point computation.


\textbf{Bitwidth Flexible Unit (BFU).} Four INT8 PEs and lightweight peripheral logic form one BFU. Following the approach of decomposing BF16 arithmetic into integer operations~\cite{wu2025tataa}, the BFU implements a four-stage pipeline (II = 1, latency = 4 cycles) supporting three BF16 operations: floating-point addition (\texttt{fpadd}), floating-point multiplication (\texttt{fpmul}), and a special operation (\texttt{fptmp}) for fast inverse square-root approximation~\cite{lomont2003fast} used in LayerNorm and Softmax. As shown in Fig.~\ref{fig:pe_design}(c), Stage~\ding{172} performs an initial integer addition for \texttt{fptmp} while other operations bypass this stage. Stage~\ding{173} concatenates the 7-bit mantissa and the 1-bit sign bit and converts them into signed 8-bit 2's-complement values via the Signed-Digit (SD) Transform, enabling INT8 PE reuse for all mantissa arithmetic. Stage~\ding{174} processes exponents through Clamp Logic or a two's-power LUT, depending on the operation type. Stage~\ding{175} performs mantissa multiplication or addition with normalization to produce the final BF16 output. The peripheral logic modules (SD Transform, Clamp Logic, and Normalization) are detailed in Fig.~\ref{fig:pe_design}(d) and add minimal area overhead relative to the INT8 PEs, completing BF16 support without dedicated floating-point hardware.

\subsection{Online Quantization and Dequantization}\label{subsec:qdq}

As described in Sec.~\ref{subsec: quantization}, VersaQ-3D's quantization flow requires precision transitions between INT and BF16 at layer boundaries. If these transitions were performed off-chip, the resulting memory traffic would negate the bandwidth savings from quantization. To keep these transitions on-chip, the accelerator includes dedicated dequantization and quantization units (lower portion of Fig.~\ref{fig:arch}) that form a pipelined precision-transition path. The dequantization unit converts INT partial-sum results from the PE array back to BF16 using pre-computed scale factors, $s_w$ for weight and $s_a$ for activations. The quantization unit then derives a per-token scale factor from the maximum absolute value across tiles and maps each BF16 element back to the target INT format via scaling, rounding, and clamping. Both units are throughput-matched to the PE array to avoid pipeline stalls, ensuring that precision transitions add negligible overhead to the end-to-end dataflow.

\subsection{Two-Stage Recomputation-Based Tiling}\label{subsec:tiling}

\begin{figure*}[t]
    \centering
    \includegraphics[width=\linewidth]{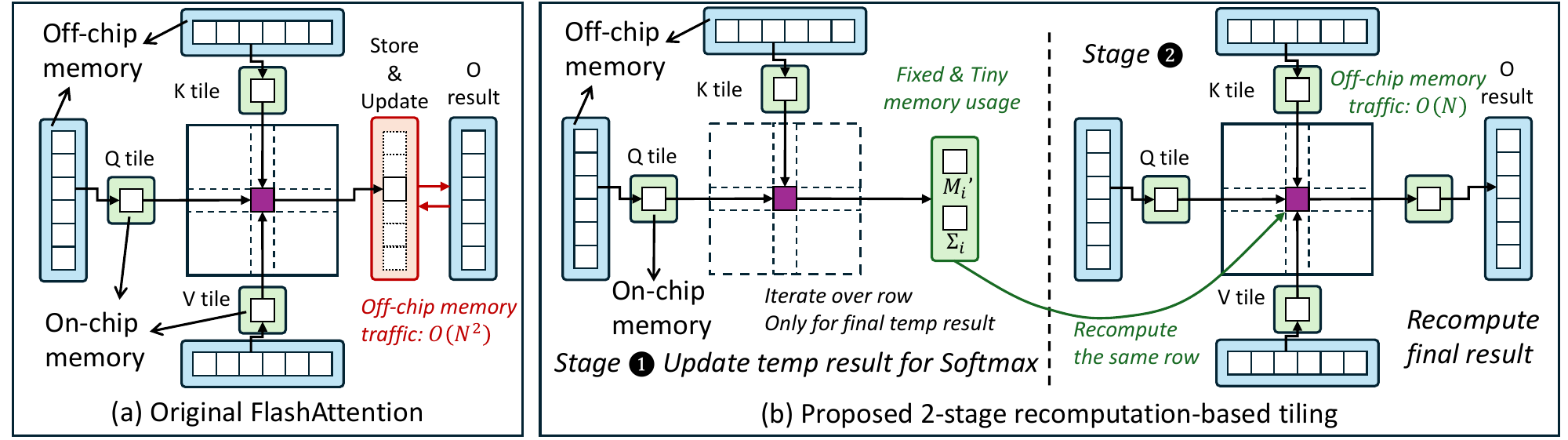}
    \caption{Two-stage recomputation-based tiling solving the memory bottleneck caused by long-sequence attention. Blue blocks for off-chip memory and green blocks for on-chip SRAM. (a) Naïvely deploying FlashAttention on our accelerator incurs $O(N^2)$ off-chip read-modify-write traffic for the O tile due to running softmax rescaling at every K tile. (b) Our method decouples statistics accumulation (Stage \ding{182}) from output computation (Stage \ding{183}), requiring only fixed and tiny on-chip storage for intermediate statistics and reducing O-tile traffic to $O(N)$.}
    \label{fig:tile}
\end{figure*}

VGGT's global attention treats all $S$ frames as a single sequence of length $N=S\times P$, producing a score matrix of size $[N,N]$ that scales quadratically with $N$ and far exceeds on-chip SRAM capacity. To tile this computation under the tight SRAM budget of our accelerator, we propose a two-stage recomputation-based tiling strategy that decouples Softmax statistics accumulation from output computation, reducing the off-chip traffic for the output matrix $O$ from $O(N^2)$ to $O(N)$.

\textbf{Limitation of FlashAttention tiling under constrained SRAM.}
FlashAttention2 (FA2)~\cite{dao2024flashattention} pairs blocked $Q$, $K$, $V$
matrix multiplication with online Softmax. Its key property is that each output
tile $O_i$ remains in on-chip SRAM for the entire attention computation. As the
$K$, $V$ tiles are scanned, $O_i$ is rescaled and accumulated in place,
\begin{equation}
  O_i^{(j)}
  = \mathrm{diag}\!\left(e^{\,M_i^{(j-1)} - M_i^{(j)}}\right)^{-1} O_i^{(j-1)}
    + \tilde{S}_i^{(j)} V_j ,
  \label{eq:fa2-update}
\end{equation}
where $M_i^{(j)}$ is the running row-wise maximum after the $j$-th $K$ tile and
$\tilde{S}_i^{(j)}$ is the corresponding score block. Since $O_i$ stays in SRAM throughout, it is written off-chip only once the entire $K$, $V$ scan completes, keeping the output traffic at $O(N)$ on GPUs.
This benefit relies on $O_i$ staying SRAM-resident for the whole computation,
which does not hold on our systolic-array-based accelerator. As discussed in
Section~\ref{subsec:challenge}, the SRAM budget is small and is already occupied
by the data movement and intermediate results of the $QK^\top$ and $SV$ matrix
multiplications~\cite{lin2025systolicattention, wang2023cosa}, leaving no room to keep $O_i$ on-chip across the full $K$, $V$
scan. As shown in Fig.~\ref{fig:tile}(a), $O_i$ must instead be read from LPDDR,
rescaled, and written back each time a new $K$, $V$ tile is processed.
Since each of the $\lceil N/T_Q \rceil$ output blocks is read and written once
per $K$ tile, the total off-chip traffic for $O$ is
\begin{equation}
  \mathcal{T}_O
  = 2 \left\lceil \tfrac{N}{T_Q} \right\rceil
       \left\lceil \tfrac{N}{T_K} \right\rceil
       T_Q \, d_k
  = O(N^{2}),
  \label{eq:traffic}
\end{equation}
where the factor of $2$ counts both the read and the write, $T_Q$ and $T_K$ are
the query and key tile sizes, and $d_k$ is the per-head dimension. In short, the
$O(N)$ traffic of FA2 stems entirely from keeping $O_i$ on-chip. Once SRAM
cannot hold it, the in-place rescaling in Eq.~\eqref{eq:fa2-update} degenerates
into repeated off-chip read-modify-writes, and the traffic grows back to
$O(N^2)$. Over the bandwidth-constrained LPDDR interface, this quadratic traffic
becomes the dominant bottleneck as $N$ grows.

\textbf{Key insight: decouple statistics from output via staged recomputation.}
Our two-stage strategy eliminates this repeated access by completing all
statistics accumulation in Stage~\ding{182} before producing any output in Stage~\ding{183}, as shown in Fig.~\ref{fig:tile}(b).
Because Stage~\ding{183} operates with finalized statistics, each O block is computed
in its correct final form and written to off-chip memory exactly once,
reducing the off-chip write volume for O to
\begin{equation}
  \mathcal{T}_O'
  = \lceil\tfrac{N}{T_Q}\rceil \times T_Q \times d_k
  = O(N)
\end{equation}
The cost is recomputing the score matrix a second time.  However, this recomputation executes in INT precision at higher throughput on our systolic array, making it far less costly than the off-chip memory traffic it eliminates.

\begin{algorithm}[t]
\caption{Proposed Tiling Method for Global Attention}
\label{alg:mha}
\begin{algorithmic}[1]
\Require $Q, K, V$, head dimension $d_k$
\Statex\hspace{-\algorithmicindent}\textbf{Tile configuration:} Q-tile ($64{\times}64$), K-tile ($64{\times}64$), V-tile ($2048{\times}64$)
\Statex\hspace{-\algorithmicindent}\textbf{Tile sizes:} $T_Q{=}64$, $T_K{=}64$, $T_V{=}2048$
\Statex\hspace{-\algorithmicindent}\textbf{Tile numbers:} $N_Q=\lceil N/T_Q\rceil$, $N_K=\lceil N/T_K\rceil$, $N_V=\lceil N/T_V\rceil$
\Statex\hspace{-\algorithmicindent}\textbf{Loop index:} $i$:Q-tile id, $j$:K-tile id, $n$:V-tile id, $m$:K-tile id within $n^{th}$ V-tile
\Statex\hspace{-\algorithmicindent}\textbf{Result:} $\mathcal{S}$: softmax result, $O$: final output result
\For{$i \gets 1$ \textbf{to} $N_Q$}
    \For{$j \gets 1$ \textbf{to} $N_K$}    \Comment{Stage \ding{182}: update temp result}
        \State $\mathcal{S}_{i,j} \gets Q_i K_j^{\top} / \sqrt{d_k}$
        \State $\mathrm{dequant}(\mathcal{S}_{i,j})$
        \State $\mathrm{update}(M_i', \Sigma_i')$
    \EndFor
    \For{$n \gets 1$ \textbf{to} $N_V$} \Comment{Stage \ding{183}: recompute final result}
        \For{$m \gets 1$ \textbf{to} ( $T_V / T_K$ )}   
            \State $\mathcal{S}_{i,n,m} \gets Q_i K_m^{\top} / \sqrt{d_k}$
            \State $\mathcal{S}_{i,n,m} \gets \exp(\mathcal{S}_{i,n,m} - M_i)\!/\Sigma_i$
            \State $\mathrm{quant}(\mathcal{S}_{i,n,m})$
        \EndFor
        \State $O_{i} \gets O_{i,n} + \mathcal{S}_{i,n} V_n$
    \EndFor
\EndFor
\end{algorithmic}

\end{algorithm}

\begin{table*}[t]
\centering
\caption{3D point map reconstruction results on 7-Scenes~\cite{shotton2013scene} and DTU~\cite{jensen2014large}.}
\resizebox{.9\textwidth}{!}{
\tiny
\setlength{\tabcolsep}{3pt}
\begin{tabular}{c|c|cccccc|cccccc}
\toprule
\multirow{3}{*}{Method} & \multirow{3}{*}{\shortstack{Bit-Width\\(W/A)}} & \multicolumn{6}{c|}{7-Scenes} & \multicolumn{6}{c}{DTU} \\
\cmidrule(l){3-8} \cmidrule(l){9-14}
 & & \multicolumn{2}{c}{Acc.$\downarrow$} & \multicolumn{2}{c}{Comp.$\downarrow$} & \multicolumn{2}{c|}{N.C.$\uparrow$} & \multicolumn{2}{c}{Acc.$\downarrow$} & \multicolumn{2}{c}{Comp.$\downarrow$} & \multicolumn{2}{c}{N.C.$\uparrow$} \\
\cmidrule(l){3-4} \cmidrule(l){5-6} \cmidrule(l){7-8} \cmidrule(l){9-10} \cmidrule(l){11-12} \cmidrule(l){13-14}
 & & Mean & Med. & Mean & Med. & Mean & Med. & Mean & Med. & Mean & Med. & Mean & Med. \\
\midrule
Full Prec. & 16/16 & 0.0438 & 0.0238 & 0.0546 & 0.0308 & 0.7341 & 0.8474 & 1.3118 & 0.7655 & 1.9189 & 1.0071 & 0.6857 & 0.7728 \\
\midrule
RTN & 4/4 & 0.0779 & 0.0549 & 0.1057 & 0.0728 & 0.6809 & 0.7736 & 3.6540 & 2.3713 & 2.2512 & 1.1532 & \cellcolor{lightgray}\textbf{0.6855} & 0.7707 \\
QuaRot & 4/4 & 0.0776 & 0.0487 & 0.1142 & 0.0712 & 0.6915 & 0.7890 & 2.8186 & 1.6209 & 2.0762 & 1.0741 & 0.6852 & 0.7710\\
ANT & 4/4 & 0.0654 & 0.0457 & 0.0905 & \cellcolor{lightgray}\textbf{0.0515} & 0.7063 & 0.8073 & 2.4306 & 1.4204 & 2.2053  & 1.1032 & 0.6821 & 0.7683 \\
Olive & 4/4 & 0.0655 & 0.0436 & 0.0906 & 0.0606 & 0.7033 & 0.8036 & 2.4639 & 1.3886 &2.1312  & 1.1295 & 0.6763 & 0.7636 \\
\textbf{Ours} & 4/4 & \cellcolor{lightgray}\textbf{0.0633} & \cellcolor{lightgray}\textbf{0.0428} & \cellcolor{lightgray}\textbf{0.0782} & 0.0537 & \cellcolor{lightgray}\textbf{0.7078} & \cellcolor{lightgray}\textbf{0.8127} & \cellcolor{lightgray}\textbf{2.4116} & \cellcolor{lightgray}\textbf{1.3756} & \cellcolor{lightgray}\textbf{2.0393} & \cellcolor{lightgray}\textbf{0.9906} & 0.6838 & \cellcolor{lightgray}\textbf{0.7721} \\
\midrule
RTN & 4/8 & 0.0463 & 0.0271 & 0.0610 & 0.0383 & 0.7282 & 0.8385 & 1.4980 & 0.8452 & 2.1069 & 1.0600 & 0.6639 & 0.7484 \\
QuaRot & 4/8 & 0.0444 & \cellcolor{lightgray}\textbf{0.0240} & 0.0559 & 0.0329 & 0.7303 & 0.8414 & 1.6475 & 0.9279 & 2.0217 & 0.9893 & 0.6678 & 0.7538 \\
ANT & 4/8 & 0.0486 & 0.0284 & 0.0585 & 0.0351 & 0.7271 & 0.8400 & 1.4597 & 0.8021 & 2.0364 & 1.0250 & 0.6582 & 0.7412 \\
Olive & 4/8 & 0.0492 & 0.0302 & 0.0804 & 0.0662 & 0.7213 & 0.8282 & 1.9331 & 1.0781 & 2.0864 & 1.0559 & 0.6739 & 0.7614 \\
\textbf{Ours} & 4/8 & \cellcolor{lightgray}\textbf{0.0438} & 0.0243 & \cellcolor{lightgray}\textbf{0.0549} & \cellcolor{lightgray}\textbf{0.0321} & \cellcolor{lightgray}\textbf{0.7343} & \cellcolor{lightgray}\textbf{0.8476} & \cellcolor{lightgray}\textbf{1.3171} & \cellcolor{lightgray}\textbf{0.7515} & \cellcolor{lightgray}\textbf{1.8805} & \cellcolor{lightgray}\textbf{0.9744} & \cellcolor{lightgray}\textbf{0.6805} & \cellcolor{lightgray}\textbf{0.7664} \\
\bottomrule
\end{tabular}
}
\label{tab:7scenes_dtu}
\end{table*}

\textbf{Two-stage execution.}
Algorithm~\ref{alg:mha} details the complete procedure.  For each Q tile
$Q_i$, Stage~\ding{182} scans all K tiles to accumulate the Softmax
statistics---the row-wise running maximum $M_i'$ and exponential sum
$\Sigma_i'$---by computing each score block in INT, dequantizing, and
applying the online update:
\begin{equation}
  M_i' = \max\bigl(M_i,\; M_{i,j}\bigr),
\end{equation}
\begin{equation}
  \Sigma_i'
  = \Sigma_i \cdot e^{(M_i - M_i')}
  + \Sigma_{i,j}\cdot e^{(M_{i,j} - M_i')},
\end{equation}
where $M_{i,j}$ is the running maximum for $j^{th}$ K tile and $\Sigma_{i,j}$ is the exponential sum for $j^{th}$ K tile.
Only the current Q tile, one K tile, and the compact statistics vectors reside
on-chip during this stage, as shown in Fig.~\ref{fig:tile}(b) green blocks.  Once the statistics
are finalized, Stage~\ding{183} recomputes the score tiles against
reorganized, larger K/V tiles ($2048\times64$) for improved data reuse and
normalizes them with the final statistics:
\begin{equation}
  \mathrm{Softmax}(x) = \frac{e^{\,x - M_i'}}{\Sigma_i'}.
\end{equation}
The normalized weights are immediately multiplied with the corresponding V
tile and used to compute the final result $O_i$ by accumulating the temporary output result and $SV$ for $n^{th}$ tile, namely $O_{i,n}$ and $S_{i,n}V_{n}$.  Because no future updates are needed, each
$O_i$ is written to off-chip memory exactly once upon completion, eliminating
both the need for an on-chip O-tile buffer and the quadratic traffic of the
naive approach.


\section{Evaluation}
\subsection{Experimental Methodology}

\textbf{Evaluation Benchmarks.} We evaluate quantization accuracy across two representative tasks and four standard 3D vision benchmarks. 
\uline{\textit{Camera Pose Estimation}} recovers the relative position and orientation between two cameras from an image pair. We use two datasets: \textbf{Co3Dv2}~\cite{reizenstein2021common}, which contains object-centric videos from the official single-sequence subset, and \textbf{RealEstate10K}~\cite{zhou2018stereo}, which covers indoor scenes. We report the Area Under the Curve (AUC) of the accuracy-threshold curve at a threshold of $30^{\circ}$. A prediction is considered correct at threshold $\tau$ only if both the relative rotation accuracy (RRA) and relative translation accuracy (RTA) fall below $\tau$. Higher AUC indicates better quality.
\uline{\textit{3D Point Map Reconstruction}} predicts dense per-pixel 3D coordinates to recover the underlying scene geometry. We use two datasets: \textbf{DTU}~\cite{jensen2014large}, a multi-view stereo benchmark with ground-truth 3D scans, and \textbf{7-Scenes}~\cite{shotton2013scene}, an RGB-D indoor benchmark with ground-truth depth and poses. We report three complementary metrics~\cite{knapitsch2017tanks}: \textit{Accuracy} measures the predicted-to-GT distance, where lower is better. \textit{Completeness} measures the GT-to-predicted distance, where lower is better. \textit{Normal Consistency} (N.C.) measures surface normal agreement, where higher is better. Since the predicted and ground-truth point clouds are normalized to a common scale before computing Accuracy and Completeness~\cite{wang2024dust3r}, the reported errors are fractions of the scene scale rather than absolute distances, leading to relatively small values and improvements.

\begin{table}[t]
\centering
\caption{Camera pose estimation results on Co3Dv2~\cite{reizenstein2021common} and RealEstate10K~\cite{zhou2018stereo}.}
\resizebox{0.48\textwidth}{!}
{
\large
\setlength{\tabcolsep}{4pt}
\begin{tabular}{c|c|ccc|ccc}
\toprule
\multirow{2}{*}{Method} & Bit-Width& \multicolumn{3}{c|}{Co3Dv2} & \multicolumn{3}{c}{RealEstate10K} \\
\cmidrule(l){3-5}\cmidrule(l){6-8}
 & (W/A) & RRA@30$\uparrow$ & RTA@30$\uparrow$ & AUC@30$\uparrow$ & RRA@30$\uparrow$ & RTA@30$\uparrow$ & AUC@30$\uparrow$ \\
\midrule
Full Prec. & 16/16 & 0.9822& 0.9941& 0.8979& 0.9994 & 0.9402 & 0.7980\\
\midrule
RTN & 4/4 & 0.9674& 0.9185& 0.6324& 0.9922 & 0.6758 & 0.4096\\
QuaRot & 4/4 & 0.9674& 0.9556& 0.6889& 0.9957 & 0.7306 & 0.4644\\
ANT & 4/4 & 0.9589 & 0.9211 & 0.6318 & 0.9986 & 0.7345 & 0.4647 \\
Olive & 4/4 & \cellcolor{lightgray}\textbf{0.9722} & 0.9267 & 0.6471 & 0.9972 & 0.7411 & 0.4691 \\
\textbf{Ours} & 4/4 & 0.9659 & \cellcolor{lightgray}\textbf{0.9615}& \cellcolor{lightgray}\textbf{0.7506}&\cellcolor{lightgray}\textbf{0.9973} & \cellcolor{lightgray}\textbf{0.7910} & \cellcolor{lightgray}\textbf{0.5371} \\
\midrule
RTN & 4/8 & 0.9719& 0.9911& 0.8888& \cellcolor{lightgray}\textbf{0.9995} & 0.9310 & 0.7728 \\
QuaRot & 4/8 & 0.9778& \cellcolor{lightgray}\textbf{0.9926}& 0.8881& 0.9993 & 0.9344 & 0.7800 \\
ANT & 4/8 & 0.9844 & 0.9922 & 0.8920 & 0.9994 & 0.9211 & 0.7489 \\
Olive & 4/8 & 0.9833 & 0.9844 & 0.8584 & 0.9991 & 0.8969 & 0.6868 \\
\textbf{Ours}& 4/8 &  \cellcolor{lightgray}\textbf{0.9837}& 0.9911& \cellcolor{lightgray}\textbf{0.8963}&0.9993 & \cellcolor{lightgray}\textbf{0.9354} & \cellcolor{lightgray}\textbf{0.7833} \\
\bottomrule
\end{tabular}
}
\label{tab:co3d}
\end{table}

\textbf{Hardware Implementation.} We implement VersaQ-3D in RTL and synthesize our design using Synopsys Design Compiler based on TSMC 28nm CMOS technology to obtain the power and area metrics. On-chip SRAMs are generated by the provided memory compiler with the same technology. We set the operating clock frequency targeting 1 GHz, which is a commonly adopted configuration for 28nm technology in prior accelerator designs~\cite{lee2023neurex, chen2025bitmod, lee2024gscore}. To evaluate the overall performance of VersaQ-3D, we develop a cycle-level simulator verified against our RTL design. The DRAM timing and power characteristics are obtained using Ramulator2~\cite{luo2023ramulator} with the configuration of LPDDR5-6400, which provides a bandwidth of 102.4 GB/s. To evaluate the quantization accuracy, we implement our algorithm
in PyTorch~\cite{paszke2019pytorch} based on the open-source code
of VGGT~\cite{wang2025vggt}. The pre-trained model we adopt is the
official VGGT-1B model from Hugging Face.

\begin{figure*}[t]
    \centering

    \subfloat[Normalized speedup over baseline devices.\label{fig:speedup}]{
        \includegraphics[width=\linewidth]{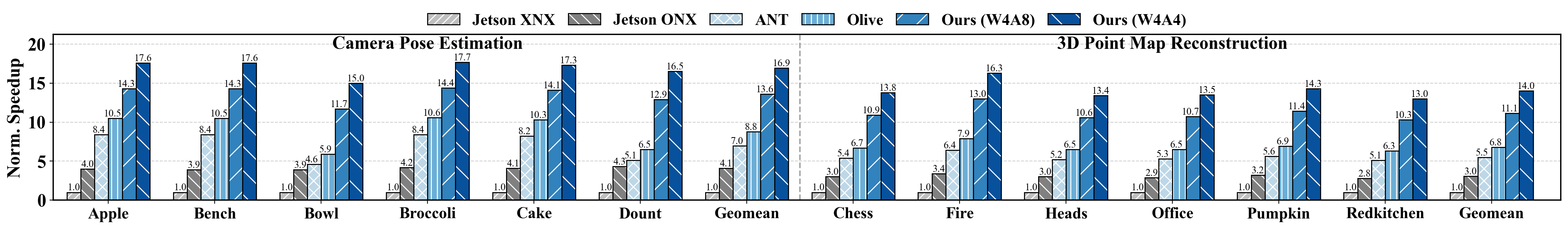}
    }

    \subfloat[Normalized energy efficiency over baseline devices.\label{fig:energy_eff}]{
        \includegraphics[width=\linewidth]{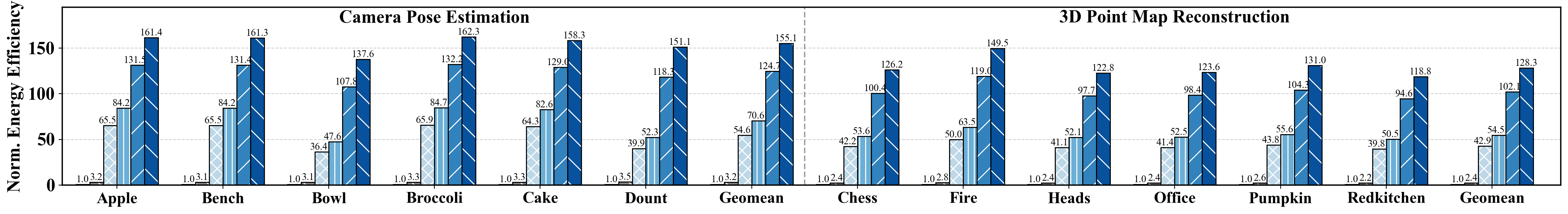}
    }

    \caption{The normalized speedup and energy efficiency achieved by our methods over baselines on Co3Dv2 and 7-Scenes.}
    \label{fig:hw_result}
\end{figure*}

\textbf{Baselines.} \uline{\textit{Quantization Baselines:}} We compare VersaQ-3D against several representative quantization methods, including both calibration-free approaches and techniques adopted by quantization-based transformer accelerators:
\begin{itemize}
    \item RTN directly quantizes weights and activations to the target bit-width using round-to-nearest.
    \item QuaRot~\cite{ashkboos2024quarot} applies Hadamard rotation to weights and activations to reduce outliers before quantization.
    \item ANT~\cite{guo2022ant} packages multiple data types and selects among them to quantize tensors at channel/token granularity.
    \item Olive~\cite{guo2023olive} adopts an outlier-victim pairing strategy to encode and preserve outlier values.
\end{itemize}
All baselines are evaluated under the same W4A8 and W4A4 configurations as our method, with quantization applied to both linear and attention layers. All methods operate at channel/token-wise granularity and quantize both weights and activations to ensure a fair comparison.

\begin{figure}[!t]
    \centering
    \includegraphics[width=\linewidth]{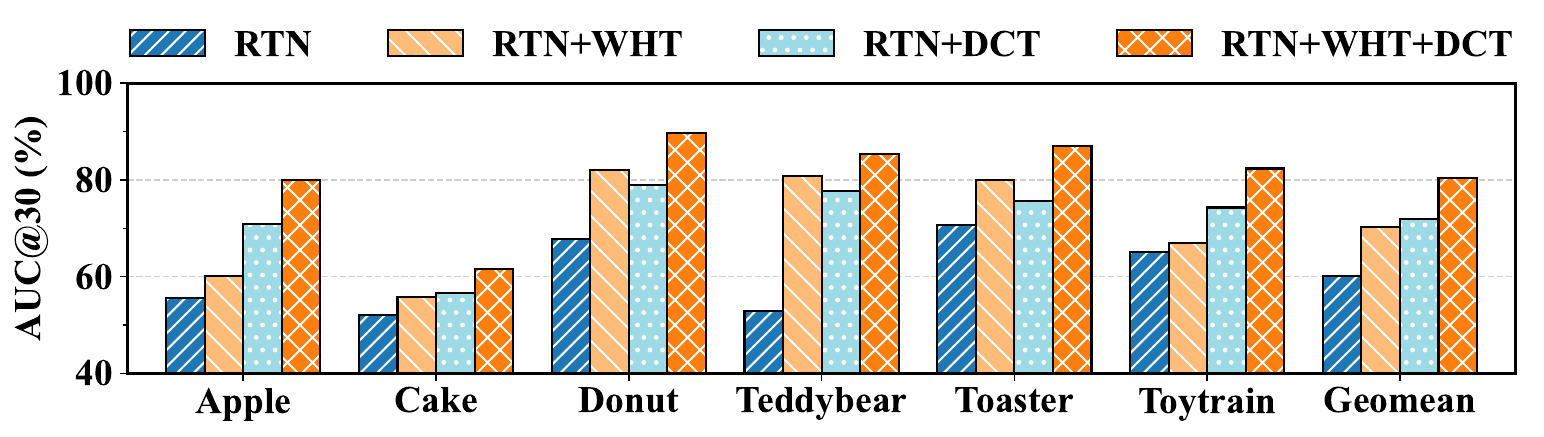}
    \caption{Ablation study of VersaQ-3D on Co3Dv2 (W4A4).}
    \label{fig:sw_ablation}
\end{figure}

\uline{\textit{Hardware Baselines:}}
To ensure a fair comparison, we consider two categories of baselines.
First, since VGGT is fundamentally built on the transformer architecture,
we select two open-source quantization-based transformer accelerators as hardware
baselines: ANT~\cite{guo2022ant} and Olive~\cite{guo2023olive}.
We follow the configurations reported in their original papers, adopting their respective 4/8-bit mixed-precision settings.
To enable a fair comparison under an iso-area constraint, we scale the PE array size of both designs so that their PE area matches that of our accelerator, with all designs synthesized under the same technology node.
Second, we include two edge GPUs widely adopted as baselines in recent 3D
reconstruction accelerator studies~\cite{li2023instant,lee2023neurex,
lee2024gscore}: Jetson Xavier NX (XNX)~\cite{ditty2018nvidia} (20\,W typical), 
Jetson Orin NX (ONX)~\cite{ditty2022nvidia} (25\,W typical), and Jetson AGX Thor~\cite{nvidia_jetson_thor_product_page} (40-130\,W typical) whose runtime
power is measured with the \texttt{jtop} power monitor.
We do not compare against prior NeRF/3DGS-based 3D vision accelerators,
as they target per-scene photorealistic view synthesis with fundamentally
different model structures and training paradigms;
Sec.~\hyperlink{par:3D}{6} provides a detailed discussion.

\subsection{VersaQ-3D Algorithm Performance}
\textbf{Quantization Accuracy.}
Tables~\ref{tab:7scenes_dtu} and~\ref{tab:co3d} summarize the
quantization accuracy of VersaQ-3D across two tasks and four
benchmarks. We compare against four baselines: RTN,
QuaRot~\cite{ashkboos2024quarot}, ANT~\cite{guo2022ant}, and
Olive~\cite{guo2023olive}.

In the aggressive W4A4 setting, VersaQ-3D consistently outperforms
all baselines on both tasks. On camera pose estimation, it achieves
an AUC@30 of 0.7506 on Co3Dv2 and 0.5371 on RealEstate10K,
surpassing the second-best method by 8.9\% and 14.5\%,
respectively. On 3D point map reconstruction, VersaQ-3D obtains
the best accuracy and completeness on both 7-Scenes and DTU while
maintaining competitive normal consistency. These
results show that meaningful 3D reconstruction remains feasible
even at aggressive 4-bit precision.

In the conservative W4A8 setting, VersaQ-3D incurs less than 1\% accuracy loss compared to full-precision results while outperforming all baselines. On Co3Dv2, the AUC@30 drops by only 0.2\%. On 7-Scenes, VersaQ-3D matches the full-precision results across all three metrics, confirming that the transform coding introduces negligible information loss at this precision. On DTU and RealEstate10K, VersaQ-3D also achieves the best performance among all quantized methods. Notably, both calibration-dependent baselines (ANT and Olive) exhibit inconsistent behavior across datasets: Olive degrades substantially on 7-Scenes W4A8 (Comp.\ 0.0804 vs.\ 0.0549 for ours), and ANT underperforms on RealEstate10K W4A8 (AUC@30 0.7489 vs.\ 0.7833 for ours). This highlights the unreliability of calibration-based approaches on diverse 3D datasets, where calibration can overfit to specific data distributions or fail to capture the full range of scene characteristics. In contrast, VersaQ-3D operates in a fully calibration-free manner without any sampled data, yet achieves robust accuracy across all datasets and tasks.

It is worth noting that there are occasionally outliers in Table~\ref{tab:7scenes_dtu}. This is because Acc., Comp., and N.C. evaluate complementary aspects of point-map quality: point accuracy, surface coverage, and normal/orientation consistency~\cite{wang2024dust3r}. Thus, a method may improve one metric while slightly underperforming on another. This accuracy–completeness trade-off is standard in 3D reconstruction evaluation, where precision/accuracy/completeness are treated as complementary criteria rather than interchangeable ones~\cite{knapitsch2017tanks}.

\begin{figure}[!t]
    \centering
    \includegraphics[width=\linewidth]{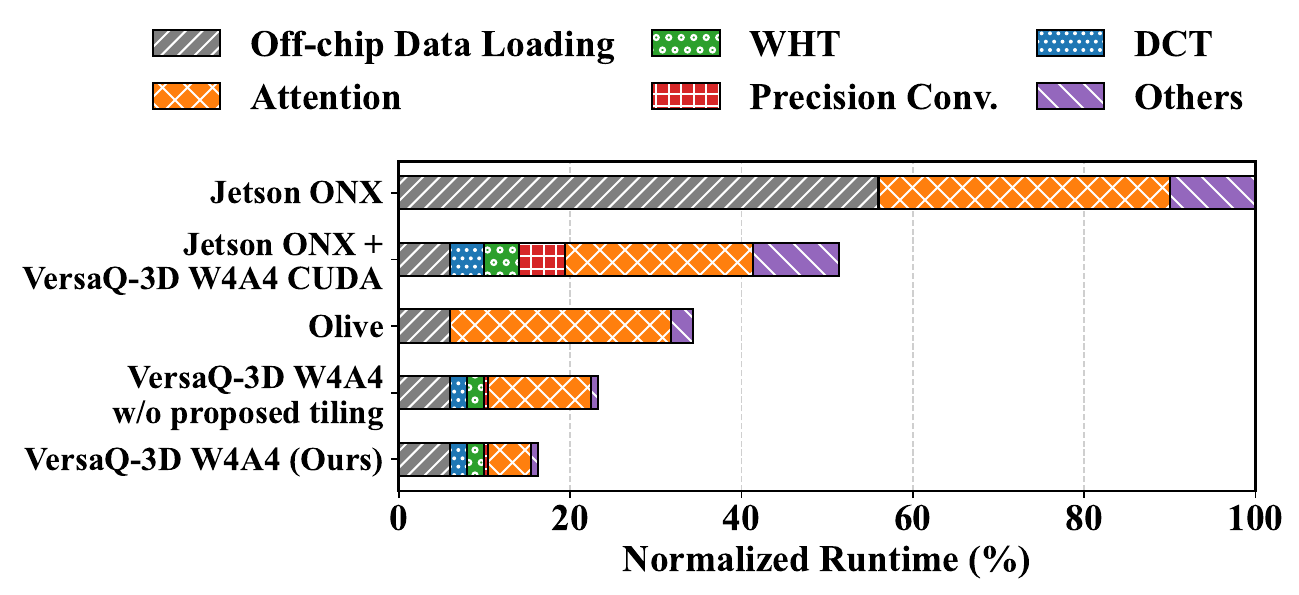}
    \caption{Runtime breakdown on Co3Dv2 (W4A4). Our quantized VGGT accelerator cuts end‑to‑end runtime by 77\% vs. Jetson ONX, and the proposed tiling scheme yields a further 7\% reduction.}
    \label{fig:runtime}
\end{figure}

\textbf{Ablation Study on Algorithm.} Fig.~\ref{fig:sw_ablation} shows the AUC@30 changes when applying each step individually in W4A4 setting on Co3Dv2 dataset. It shows that applying WHT and DCT individually leads to a moderate accuracy increase. This is because the WHT redistributes the outliers and produces a more uniform and quantization-friendly distribution for activations. However, the accuracy after only adopting WHT is still not satisfactory. It is interesting to note that on Apple and Toytrain datasets, WHT only brings a marginal increase in the AUC@30. This further justifies the limitation of WHT and the necessity to adopt further steps. In contrast, the DCT-only method lacks the capability to amortize the outliers, leading to sub-optimal accuracy. Therefore, we further combine the DCT with WHT to maintain the structural features of weights to mitigate the quantization errors. The results indicate that with each step, the accuracy increases by 10\% on average, respectively, validating the effectiveness of the proposed VersaQ-3D.

\subsection{VersaQ-3D Hardware Performance}

\textbf{Area and Power.} Table~\ref{tab:area_power_breakdown} summarizes the hardware utilization and implementation results of VersaQ-3D. The accelerator integrates 64$\times$16 BFUs, which is equivalent to 64$\times$64 INT8 PEs and 128$\times$128 INT4 PEs under their respective modes. VersaQ-3D occupies 4.17 mm$^2$ of area and consumes 2.17 W of typical power. Table~\ref{tab:area_power_breakdown} also presents the detailed area and power breakdown, including per‑BFU, per‑INT8‑PE, and per‑INT4‑PE statistics, showing the contribution of each architectural component.


\textbf{Performance.} Fig.~\ref{fig:hw_result} shows the speedup and energy efficiency of VersaQ‑3D over the baseline devices. We additionally compare against two quantization‑based accelerators, ANT~\cite{guo2022ant} and Olive~\cite{guo2023olive}. Under the W4A4 setting targeting instant 3D reconstruction, VersaQ‑3D achieves 22.0$\times$ and 5.4$\times$ average speedup over Jetson XNX and ONX on camera pose estimation (Co3Dv2), and 17.9$\times$ and 6.0$\times$ on 3D point‑map reconstruction (7‑Scenes). Compared with ANT and Olive, VersaQ‑3D delivers 2.9$\times$/2.2$\times$ speedup on Co3Dv2 and 3.0$\times$/2.4$\times$ on 7‑Scenes, respectively. These gains stem from reduced memory traffic and the lower computational complexity enabled by INT4 quantization. For energy efficiency, VersaQ‑3D achieves 202.0$\times$ and 63.1$\times$ gains over Jetson XNX and ONX on Co3Dv2, and 164.7$\times$ and 68.6$\times$ on 7‑Scenes. Compared with ANT and Olive, VersaQ‑3D outperforms by 3.3$\times$/2.5$\times$ on Co3Dv2 and 3.5$\times$/2.7$\times$ on 7‑Scenes, respectively. VersaQ‑3D in W4A8 mode, which targets high‑fidelity reconstruction, also consistently outperforms all baselines, further demonstrating the efficiency and versatility of our design.

\begin{figure}[!t]
    \centering
    \includegraphics[width=\linewidth]{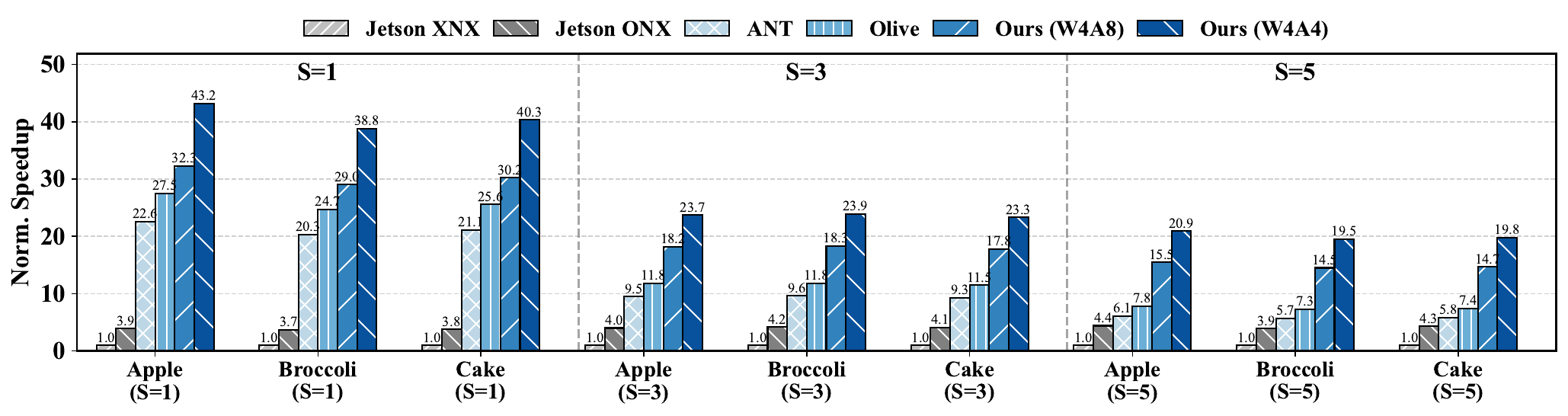}
    \caption{Normalized speedup vs. frame count $S$ on Co3Dv2.}
    \label{fig:robustness}
\end{figure}

\textbf{Ablation Study and Robustness Analysis.} To quantify the contribution of each design component, we perform an ablation study by incrementally applying the proposed quantization and tiling techniques. The corresponding runtime breakdown is shown in Fig.~\ref{fig:runtime}. Quantization substantially reduces off‑chip data loading time, while the attention stage remains a dominant latency source. The same effect appears in Olive, where non-linear operations are offloaded to a dedicated FP unit and long global-attention sequences incur heavy off-chip traffic, both lengthening attention latency. Adding the proposed tiling strategy lowers attention runtime by 7\% over VersaQ-3D without tiling, confirming its effectiveness. We also implement our algorithm in CUDA W4A4 on Jetson ONX, which does not reach the theoretical INT4 Tensor-Core speedup end-to-end due to substantial non-Matrix Multiply Accumulate (MMA) work. In contrast, our on-chip online quantization reduces precision-conversion time by 92\% over this baseline and adds only 4\% runtime over the original VGGT model, yielding an overall 5.3$\times$ speedup over Jetson ONX, mainly from reduced weight loading (32-bit $\rightarrow$ 4-bit) and 4-bit computation. We further evaluate the robustness of VersaQ-3D under different frame numbers ($S$), as illustrated in Fig.~\ref{fig:robustness}. The results show a consistent speedup over baseline designs across all settings. Notably, the speedup is most pronounced when 
$S=1$, where the memory bottleneck dominates and the reduction in off‑chip data loading yields a larger overall performance gain.

\begin{figure}
    \centering
    \includegraphics[width=\linewidth]{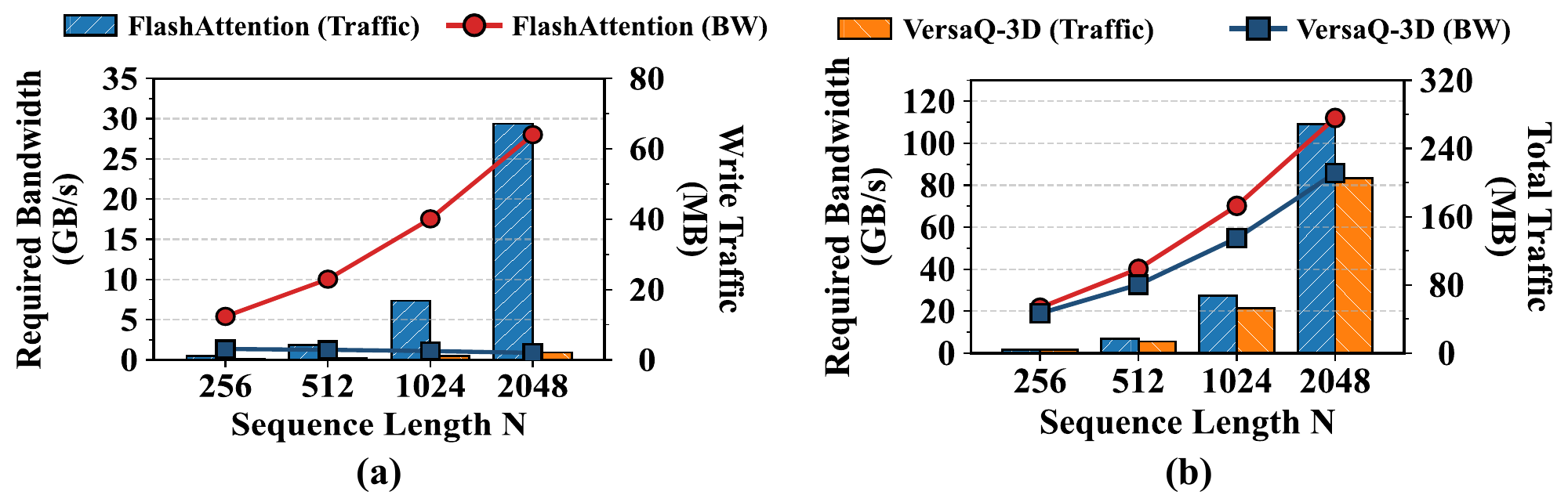}
    \caption{Off-chip memory traffic vs. sequence length for FlashAttention~\cite{dao2024flashattention} and VersaQ-3D (Ours). (a) Off-chip write traffic. (b) Total off-chip traffic.}
    \label{fig:mem}
\end{figure}

\begin{table}[ht]
\centering
\caption{Area/Performance tradeoff between our BFU and heterogeneous PE under same equivalent throughput.}
\label{tab:area_perf_tradeoff}
\resizebox{0.48\textwidth}{!}{
\begin{tabular}{lcccc}
\hline
\textbf{Type} & 
\makecell{\textbf{INT4 OPS/Area}\\ \textbf{(TOPS/mm$^2$)}} & 
\makecell{\textbf{INT8 OPS/Area}\\ \textbf{(TOPS/mm$^2$)}} & 
\makecell{\textbf{BF16 OPS/Area}\\ \textbf{(GOPS/mm$^2$)}} & 
\makecell{\textbf{Latency/Area}\\ \textbf{(ms/mm$^2$)}} \\
\hline
\large BFU  & \large 5.9  & \large 1.5  & \large 92  & \large 166 \\
\large INT4+BF16   & \large 5.3  & \large N.A. & \large 83  & \large 174 \\
\large INT8+BF16   & \large N.A. & \large 2.3  & \large 142 & \large 449 \\
\hline
\end{tabular}
}
\end{table}

\textbf{Off-Chip Traffic and Tradeoff Analysis.} To illustrate the off-chip memory traffic reduction, we compare the write and total off-chip traffic volume and bandwidth between FlashAttention2~\cite{dao2024flashattention} and VersaQ-3D in Fig.~\ref{fig:mem}. As shown in Fig.~\ref{fig:mem}(a), when the sequence length \(N\) increases from 256 to 2048, the write traffic of FlashAttention2 grows from 1~MB to 67~MB and its bandwidth increases from 5~GB/s to 28~GB/s, while VersaQ-3D only increases from 0.2~MB to 2~MB and keeps the write bandwidth low at 1~GB/s. This further validates our two-stage recomputation-based tiling method reduces the write traffic into $O(N)$. Fig.~\ref{fig:mem}(b) shows a similar trend in total off-chip traffic: FlashAttention2 requires 4$\to$268~MB and 21$\to$112~GB/s, whereas VersaQ-3D reduces them to 3$\to$205~MB and 18$\to$85~GB/s. We further evaluate the area/performance tradeoff under the same equivalent throughput for the supported precisions in Table~\ref{tab:area_perf_tradeoff}. The proposed BFU achieves the best latency/area efficiency, 166~ms/mm\(^2\), compared with 174~ms/mm\(^2\) for INT4+BF16 and 449~ms/mm\(^2\) for INT8+BF16. Although INT8+BF16 occasionally shows higher OPS/area due to its specialized datapath and the absence of BFU reconfiguration overhead, it cannot support INT4 and its peak compute density cannot be fully translated into lower latency because of workload imbalance. 

\section{Related Work}\label{sec:related}
\textbf{Mixed-precision accelerators.} Prior work on mixed-precision accelerators can be grouped into three categories. First, INT mix-precision designs, such as BitFusion~\cite{sharma2018bit} and BitBlade~\cite{ryu2019bitblade}, flexibly support different integer bit widths, but they only target INT computation and lack support for FP non-linear operators. Second, weight-only quantization designs, such as FIGNA~\cite{jang2024figna} and BitMoD~\cite{chen2025bitmod}, use FP-INT PEs to support quantized weights, but leave activations in FP. This is not sufficient for VGGT, where the large number of image tokens makes activation quantization important and weight-only quantization inefficient. Third, INT-FP mix-precision designs, such as RaPiD~\cite{venkataramani2021rapid}, fuse FP and INT computation into a single MPE. However, RaPiD targets general DNN inference and training, and does not optimize attention-specific computation.

\textbf{Long-context/sequence solutions.} Prior work on long-context and long-sequence processing can be grouped into three directions. First, tiling-based attention designs such as FLAT~\cite{kao2023flat} and FuseMax~\cite{nayak2024fusemax} reduce memory traffic through fusion and data reuse. FLAT targets the memory-bandwidth bottleneck on spatial accelerators, but it needs to store the full softmax vector, causing large memory overhead as the sequence length grows. FuseMax reduces this bottleneck with a one-pass attention cascade, but it still keeps many high-precision intermediate results on chip for updates, which exceeds our SRAM budget. Second, KV-cache management methods such as Oaken~\cite{kim2025oaken} and V-Rex~\cite{kim2026v} are mainly designed for autoregressive LLM decoding, where past tokens are reused in later decoding steps. In contrast, VGGT performs global attention over concatenated tokens from all frames at once, so these KV-cache methods cannot be directly applied. Third, SSM-based or global-convolution models such as VGA~\cite{lee2024vga} use convolution as a scalable alternative to self-attention. However, adopting these methods would require major changes to the VGGT architecture and costly retraining or fine-tuning for 3D vision tasks.

\begin{table}[!t]
\centering
\caption{Area and Power Breakdown of the Hardware Design}
\resizebox{0.48\textwidth}{!}{
\begin{tabular}{l l c c}
\toprule
\textbf{Component} & \textbf{Setup} & \textbf{Area~[mm$^2$]} & \textbf{Power~[W]} \\
\midrule
BFU Units & $64 \times 16$ (BFUs + BFU Buffer) & 2.77 & 1.79 \\
{\small \quad BFU} & {\small 4 INT8 PEs + Peripheral Logic} & {\small 2.79E-3} & {\small 1.75E-3} \\
{\small \quad \quad INT8 PE} & {\small 4 INT4 PEs + Adder} & {\small 5.95E-4} & {\small 2.32E-4} \\
{\small \quad \quad \quad INT4 PE} & {\small Basic Units} & {\small 1.04E-4} & {\small 3.69E-5} \\
Quant Unit & {\small 128 Scale and ABS Value Lanes + Comparator} & 0.04 & 0.04 \\
Dequant Unit & {\small 128 Scale + Clamp \& Round Lanes} & 0.04 & 0.02 \\
\midrule
Weight Buffer & $2 \times 128\,\mathrm{KB}$ & 0.54 & 0.13 \\
Input Buffer & $128\,\mathrm{KB}$ & 0.24 & 0.06 \\
Output Buffer & $256\,\mathrm{KB}$ & 0.54 & 0.13 \\
\midrule
\textbf{Total} &   & \textbf{4.17} & \textbf{2.17} \\
\bottomrule
\end{tabular}
}
\label{tab:area_power_breakdown}
\end{table}

\textbf{Model Quantization.} Numerous PTQ methods have been developed to reduce the memory and computation overhead of large models, including LLMs~\cite{zhang2025sageattention, xiao2023smoothquant, ashkboos2024quarot, liu2024spinquant, lin2024awq,lee2024tender} and VLMs~\cite{zhao2024vidit, zhao2024mixdq, liu2024hq, li2025mbq, federici2025hadanorm}. Most LLM quantization approaches~\cite{dettmers2022gpt3, xiao2023smoothquant, lin2024awq} rely on calibration datasets to capture data distributions, which is impractical for VGGT due to the diversity and complexity of 3D datasets. On the other hand, VLM‑oriented methods~\cite{zhao2024vidit, zhao2024mixdq, liu2024hq} mainly study the quantization effects on video generation tasks and thus are not directly applicable to VGGT. Recent work~\cite{xu2025llm} demonstrates that versatile compression of LLMs can be achieved without calibration data. Building on this insight, we propose VersaQ‑3D, a versatile quantization framework that alleviates the challenge of constructing 3D calibration datasets for VGGT.


\hypertarget{par:3D}{}\textbf{3D Graphics Accelerators.} Recent 3D graphics accelerators target either NeRF‑based~\cite{li2023instant, lee2023neurex, li2022rt, feng2024cicero} or 3DGS-based~\cite{lee2024gscore, lin2025metasapiens, feng2025lumina} novel view synthesis. We do not directly compare against them, as their objective differs fundamentally from VGGT's multi‑task 3D reconstruction (point maps, camera parameters, etc.), and their smaller, non‑transformer, per‑scene architectures make performance comparisons not meaningful.

\section{Conclusion}
In this paper, we propose VersaQ-3D, an algorithm-architecture co-design framework to enable instant, feed-forward 3D reconstruction on edge devices. On the algorithm side, VersaQ-3D employs transform-coding-based versatile quantization without requiring any calibration datasets, achieving negligible accuracy loss at W4A8 and state-of-the-art accuracy at W4A4 over prior quantization methods across four benchmarks. On the architecture side, a reconfigurable accelerator supports multiple precisions and non-linear operations and is combined with a recomputation-based tiling method to alleviate the long-sequence bottleneck, achieving significant speedup over both edge GPUs and prior quantization-based accelerators in a compact, power-efficient design. 
\bibliographystyle{ACM-Reference-Format}
\bibliography{sample-base}
\end{sloppypar}
\end{document}